\documentclass[journal=jacsat,manuscript=article]{achemso}

\usepackage{chemformula} % Formula subscripts using \ch{}
\usepackage[T1]{fontenc} % Use modern font encodings
\usepackage{amsmath,amssymb}
\author{Jiarul Midya}
\email{j.midya@fz-juelich.de}
\author{Thorsten Auth}
\email{t.auth@fz-juelich.de}
\author{Gerhard Gompper}
\affiliation{Theoretical Physics of Living Matter, Institute for Biological Information Processing and Institute for Advanced Simulation, j.midya@ J\"{u}lich,  52425 J\"{u}lich, Germany.}
\email{g.gompper@fz-juelich.de}
\title[An \textsf{achemso} demo]
{Membrane-Mediated Interactions Between Nonspherical Elastic Particles}

\def \figwidth {16cm}
\begin{document}


%\makeatletter
%\setlength\acs@tocentry@height{3.8cm}
%\setlength\acs@tocentry@width{9.0cm}
%\begin{tocentry}
%  \begin{center}
%    \includegraphics[width=8.5cm]{Figures/f-TOC.eps}
%  \end{center}
%\end{tocentry}
\begin{abstract}
The transport of particles across lipid-bilayer membranes is important for biological cells to exchange information and material with their environment. Large particles often get wrapped by membranes, a process which has been intensively investigated in the case of hard particles. However, many particles in vivo and in vitro are deformable, e.g., vesicles, filamentous viruses, macromolecular condensates, polymer-grafted nanoparticles, and microgels. Vesicles may serve as a generic model system for deformable particles. Here, we study non-spherical vesicles with various sizes, shapes, and elastic properties at initially planar lipid-bilayer membranes. Using the Helfrich  Hamiltonian, triangulated membranes, and energy minimization, we predict the interplay of vesicle shapes and wrapping states. Increasing particle softness enhances the stability of shallow-wrapped and deep-wrapped states over non-wrapped and complete-wrapped states. The free membrane mediates an interaction between partial-wrapped vesicles. For the pair interaction between deep-wrapped vesicles, we predict repulsion. For shallow-wrapped vesicles, we predict attraction for tip-to-tip orientation and repulsion for side-by-side orientation. Our predictions may guide the design and fabrication of deformable particles for efficient use in medical applications, such as targeted drug delivery.
\end{abstract}

%\section{Introduction}
Wrapping processes of mesoscopic particles by cell membranes are abundant in nature, such as entry and exit of viruses and parasites into host cells \cite{dasgupta_membrane-wrapping_2014,dasgupta_nano-_2017}. Furthermore, extracellular vesicles can be taken up via endocytosis \cite{joshi_endocytosis_2020}. Therefore, particle wrapping by lipid-bilayer membranes is also important for the design diagnostic and therapeutic agents. Particles for medical applications include lipid particles for targeted drug delivery \cite{sebastiani_apolipoprotein_2021,pardi_expression_2015}, and magnetic nanoparticles that serve as heat sources for cancer therapy \cite{maier-hauff_efficacy_2011}. For hard particles the desired wrapping states can be achieved by controlling their shape, size, and particle-membrane adhesion strength \cite{dasgupta_shape_2014}. For soft particles, in addition, particle deformability plays an important role in the wrapping process. For example, an increased stability of partial-wrapped states has been reported for initially spherical vesicles with low bending rigidities of their membrane \cite{yi_cellular_2011}. Vesicles flatten upon binding to planar substrates \cite{smith_effective_2005}, the increased stability of partial-wrapped states is therefore similar to the increased stability reported for hard oblate ellipsoidal compared with spherical particles \cite{dasgupta_wrapping_2013}. Furthermore, soft particles can adjust to constraints and react to external stimuli. Prominent examples are filamentous viruses at plasma membranes that bend \cite{welsch_electron_2010}, parasites that squeeze through narrow constrictions when invading host cells through the tight junction \cite{del_rosario_apicomplexan_2019}, and SARS-CoV-2 virions, which are bounded by a lipid envelope, and assume non-spherical shapes near membranes \cite{klein_sars-cov-2_2020}.

Engineered soft particles encompass a wide variety of architectures with tunable elastic properties: microgels \cite{wang_assembling_2019}, star polymers \cite{likos_star_1998}, polymer-grafted nanoparticles \cite{midya_structure_2020}, polymeric shells \cite{donath_novel_1998}, particles made from dendrimers as building blocks \cite{tomalia_new_1985}, vesicles \cite{seifert_configurations_1997}, and biomolecular condensates \cite{banani_biomolecular_2017}. 
For example, the elastic properties of microgels can be controlled by crosslinker density and electric charge \cite{gnan_silico_2017,hofzumahaus_monte_2021}, of star polymers by functionality and chain length \cite{daoud_star_1982}, of polymer-grafted nanoparticles by grafting density and chain length \cite{midya_structure_2020}, and of unilamellar fluid vesicles by membrane bending rigidity and osmotic concentrations \cite{yu_osmotic_2020}. Vesicles, in particular, are a versatile and well-established biomimetic system and have also been used as generic model system for soft particles \cite{dimova_giant_2019}. A zoo of vesicle shapes can be obtained by changing the membrane curvature-elastic parameters and osmotic concentrations, which includes spherical, prolate, oblate, stomatocyte, pear-shaped, and starfish shapes \cite{seifert_configurations_1997}. 

Particles have been shown to laterally move on lipid membranes in experiments using viruses on plasma membranes and synthetic particles on free-standing lipid bilayers, as well as in computer simulations \cite{kukura_high-speed_2009,sarfati_long-range_2016,edebets_characterising_2020}. Therefore, long-ranged membrane-mediated interactions can drive self-assembly of partial-wrapped particles at lipid-bilayer membranes \cite{auth_budding_2009,reynwar_aggregation_2007}. However, the nature of the membrane-mediated interactions can either be attractive or repulsive, depends on many parameters, and is thus not easy to predict. For example, both, the deformation energy of the free membrane and the particle-membrane adhesion energy depend on the distance between interacting particles \cite{saric_fluid_2012}. Experimental and theoretical studies show membrane-mediated attraction and cluster formation of hard spherical particles on vesicles \cite{idema2019interactions}. 
%Also, pair interactions between strongly curved spherical caps and many-body interactions between weakly curved inclusions have been found to be attractive \cite{reynwar_membrane-mediated_2011,kim_curvature-mediated_1998}.
For example, hard spherical particles on giant unilamellar vesicles have been found to attract each other and to induce tube formation \cite{van_der_wel_lipid_2016,saric_fluid_2012,saric_mechanism_2012,acosta-gutierrez_role_2021,bahrami_tubulation_2012}. In vivo, viruses are shed as multi-virion clusters in vesicles \cite{santiana_vesicle-cloaked_2018,sanjuan_why_2019}.
Ellipsoidal microgels on giant unilamellar vesicles have been found to form ordered structures with spacings that suggest membrane-mediated mutual repulsion \cite{wang_assembling_2019}. %Furthermore, pair interactions of conical proteins and weakly curved spherical-cap inclusions on planar membranes are usually repulsive \cite{weikl_interaction_1998,goulian_long-range_1993}.

In this work, we use energy minimization to calculate and predict shapes and wrapping states for single, non-spherical vesicles that get wrapped at planar membranes. We find that adhesion to a membrane can change the shape of vesicles from prolate to oblate for partial-wrapped states. Wrapping transitions can be continuous and discontinuous; shape and orientation changes of vesicles are always discontinuous with an energy barrier. Increased vesicle softness, decreased reduced vesicle volume, and increased tension of the planar membrane enhance the stability of the partial-wrapped states. We also calculate the membrane-mediated pair interaction between two prolate vesicles in shallow and deep-wrapped states. We predict attraction between shallow-wrapped vesicles in tip-to-tip orientation, and repulsion between deep-wrapped vesicles, as well as between shallow-wrapped vesicles in side-by-side orientation. For shallow-wrapped vesicles, increasing particle deformability induce a shape change from prolate to oblate and an interaction change from attractive in tip-to-tip orientation to repulsive for oblate vesicles.

%We begin by discussing the wrapping energy contributions and relevant parameters. Vesicles shapes, energies and phases are predicted for single-vesicle systems. Finally, we calculate the membrane-mediated pair interaction between partial-wrapped vesicles.

\section{Results and Discussion}

{\bf Wrapping energies.} Our predictions for wrapping vesicles at membranes are based on a continuum membrane model where the membranes are represented by mathematical surfaces, which is applicable for particles with radii $r_{\rm p} \gtrsim 20 \, \rm nm$. The total energy of the system is calculated using the energy functional
\begin{equation}
E=\int_{A_{\rm p}} dS \, [2 \kappa_{\rm p} H^2 + \sigma] +  \int_{A_{\rm v}} dS \, 2 \kappa_{\rm v} H^2 - w \int_{A_{\rm ad}} dS \, ,
\label{Eq:Helfrich_Hamiltonian}
\end{equation}
where $H=(c_1+c_2)/2$ is the mean curvature and $c_1$ and $c_2$ are the two principle curvatures; $A_{\rm v}$ is the area of the vesicle membrane, $A_{\rm p}$ the area of the initially planar membrane, and $A_{\rm ad}$ is the adhered area of the vesicle membrane. The bending rigidities are $\kappa_{\rm p}$ and $\kappa_{\rm v}$ for the planar and the vesicle membrane, respectively; $\sigma$ is the tension of the planar membrane, and $w$ is the adhesion strength between vesicle and membrane. The shape of the vesicle is characterized by its reduced volume $v=V_{\rm v}/V_{\rm sph}$, where $V_{\rm v}$ is the actual volume of the vesicle and $V_{\rm sph}$ is the volume of a spherical vesicle with the same membrane area. The deformability of the vesicle is characterized by the ratio $\kappa_{\rm v}/\kappa_{\rm p}$ of the bending rigidities of vesicle and planar membrane. The wrapping fraction $f_w = A_{\rm ad}/A_{\rm v}$ measures the fraction of the vesicle membrane that is adhered. For convenience, we will use the reduced membrane tension $\tilde{\sigma} = \sigma A_{\rm v}/(2\pi\kappa_{\rm p})$ and the reduced adhesion strength $\tilde{w}=wA_{\rm v}/(\pi\kappa_{\rm p})$ in the following.

\begin{figure}[htbp!]
    \centering
    \includegraphics[width=\figwidth]{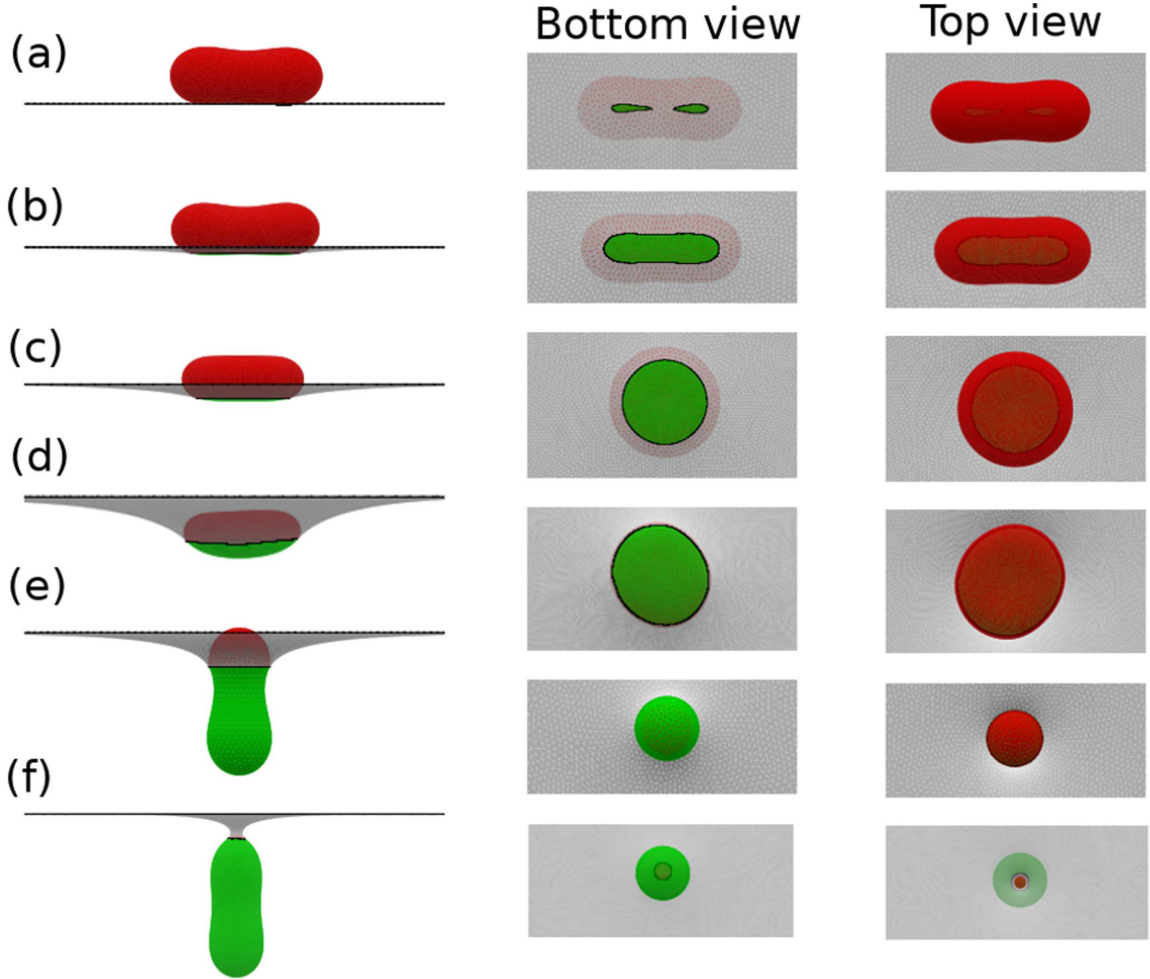}
    \caption{Wrapping of a non-spherical vesicle with $v=0.8$ at an initially planar membrane for the wrapping fractions (a) $f_w=0.05$ (unstable state), (b) $f_w=0.25$ (stable state), (c) $f_w=0.35$ (stable state), (d) $f_w=0.5$ (unstable state), (e) $f_w=0.75$ (stable state), and (f) $f_w=0.95$ (stable state). The minimum-energy shape of the free vesicle is prolate. The bending rigidity ratio is $\kappa_{\rm v}/\kappa_{\rm p}=1$ and the reduced membrane tension $\tilde{\sigma}=0.5$. The left, middle, and right columns represent the side, bottom, and top views of the vesicle-membrane system, respectively.}
    \label{fig:snapshots}
\end{figure}

{\bf Wrapping and shape transitions for a non-spherical vesicle.} The wrapping of non-spherical vesicles at a planar membrane is studied by systematically varying the bending rigidity ratio $\kappa_{\rm v}/\kappa_{\rm p}$ and reduced volume $v$ of the vesicle, and the tension $\sigma$ of the planar membrane. Figure \ref{fig:snapshots} shows minimum energy shapes of a vesicle with reduced volume $v=0.8$, at a planar membrane as function of the wrapping fraction $f_w=A_{\rm ad}/A_{\rm v}$. Wrapping always starts from the lowest mean curvature region of the vesicle surface; for a prolate vesicle binding occurs in submarine orientation with its major axis parallel to the membrane, see Fig.~\ref{fig:snapshots} (a). As wrapping progresses, the deformation of vesicle and surrounding membrane increases, see Fig.~\ref{fig:snapshots}(b)-(d). We observe a stable oblate state of the vesicle at $f_w \approx 0.3$. The deformation of the vesicles reaches its maximum at $f_w \approx 0.5$, where the vesicle shape is oblate and the membrane touches the rim of the vesicle where the mean curvature is maximal. For $f_w > 0.5$, the shape of the vesicle changes back from oblate to prolate--now with rocket orientation and its major axis perpendicular to the membrane, see Fig.~\ref{fig:snapshots}(e,f). 

\begin{figure}[htbp!]
    \centering
    \includegraphics[width=\figwidth]{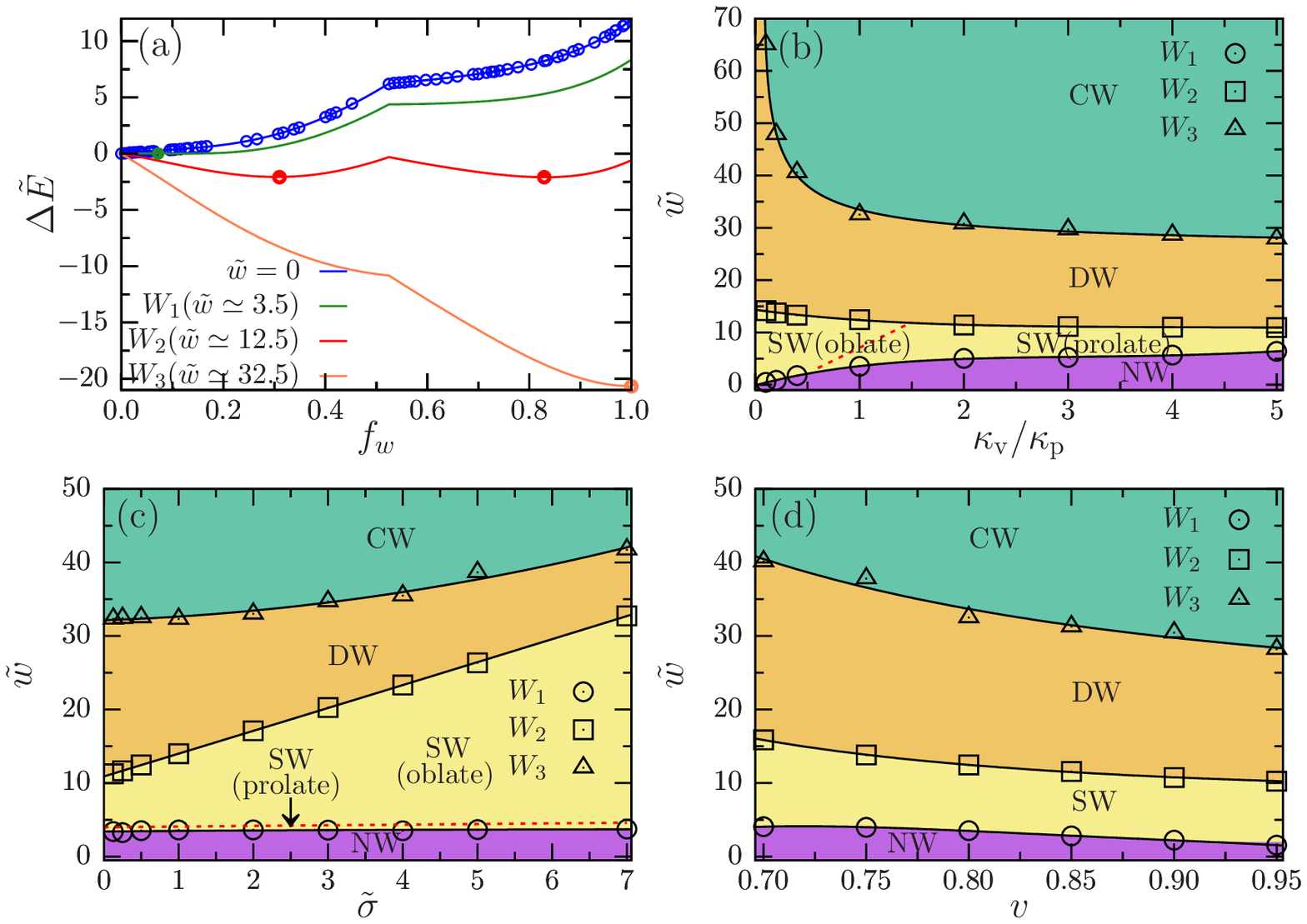}
    \caption{Wrapping of a non-spherical vesicle at a planar membrane. (a) Reduced wrapping energy $\Delta \tilde{E}$ as function of the wrapping fraction $f_w$ for vesicles with reduced volume $v=0.8$, bending rigidity ratio $\kappa_{\rm v}/\kappa_{\rm p}=5$, and reduced membrane tension $\tilde{\sigma}=0.5$. The blue circles show the numerical data for the wrapping energy in absence of adhesion strength ($\tilde{w}=0$) and the blue line represents the fit function. The green, red, and orange lines correspond to the adhesion strengths $\tilde{w}$ associated with the wrapping transitions from non-wrapped (NW) to shallow-wrapped (SW), shallow-wrapped to deep-wrapped (DW), and deep-wrapped to complete-wrapped (CW), respectively. The energies for the (coexisting) stable states at the transitions are indicated by circles. (b) Wrapping diagram in the $\kappa_{\rm v}/\kappa_{\rm p}$-$\tilde{w}$ plane for $v=0.8$ and $\tilde{\sigma}=0.5$. The dashed red line of the oblate-to-prolate transition is a guide to the eye. (c)  Wrapping diagram in the $v$-$\tilde{w}$ plane for $\kappa_{\rm v}/\kappa_{\rm p}=1$ and $\tilde{\sigma}=0.5$. (d) Wrapping diagram in the $\tilde{\sigma}$-$\tilde{w}$ plane for $v=0.8$ and $\kappa_{\rm v}/\kappa_{\rm p}=1$. All transitions and the stable wrapping states of the vesicles are labeled in the figures.}
    \label{fig:dEng+phaseDia}
\end{figure}

In absence of adhesion energy ($\tilde{w}=0$), a monotonic increase of the wrapping energy $\Delta \tilde{E}$ is observed with increasing wrapping fraction $f_w$, see Fig.~\ref{fig:dEng+phaseDia} (a). At finite adhesion strengths, depending on the elastic properties, three or four transitions between non-wrapped, partial-wrapped, and complete-wrapped states are found. For adhesion strengths below the binding transition, $\tilde{w} < \tilde{w}_{1}$, the non-wrapped state is stable. The binding transition $W_1$ at adhesion strength $w_1$ is discontinuous, see Fig.~6S in the Supporting Information, corresponding to the merging of two separate adhesion patches, see Fig.~\ref{fig:snapshots}. At higher adhesion strength, a transition from shallow-wrapped prolate in submarine orientation to shallow-wrapped oblate may occur. A further increase of the adhesion strength yields a coexistence of a shallow-wrapped state with low wrapping fraction and a deep-wrapped state of a prolate in rocket orientation with high wrapping fraction; this $W_2$ transition at adhesion strength $w_2$ is also discontinuous. The envelopment transition $W_3$ at adhesion strength $w_3$, between a deep-wrapped and the complete-wrapped state, is continuous. For adhesion strengths above the envelopment transition, $\tilde{w} \geq \tilde{w}_{3}$, the complete-wrapped state is stable. 

The wrapping diagrams of the vesicle-membrane system are obtained by systematic variation of the bending-rigidity ratio $\kappa_{\rm v}/\kappa_{p}$ that effectively controls the vesicle softness, the reduced volume $v$ that controls the shape of free vesicles, see Fig.~1S in Supporting Information, and the reduced tension $\tilde{\sigma}$ of the planar membrane. For each parameter set, we compute the deformation energy $\Delta \tilde{E}$ as function of the wrapping fraction $f_w$, see Fig.~3S-5S in the Supporting Information. Figure~\ref{fig:dEng+phaseDia}(b) shows the wrapping diagram in the $\tilde{w}$-$\kappa_{\rm v}/\kappa_{p}$ plane for fixed $v=0.8$ and $\tilde{\sigma}=0.5$. As we decrease the value of $\kappa_{\rm v}/\kappa_{p}$, the binding transition $W_1$ occurs at lower adhesion strength $\tilde{w}$; in contrast, the adhesion strength for the envelopment transition increases. Therefore, as the vesicle becomes softer, binding to the planar membrane becomes easier and complete wrapping more difficult. For the $W_2$ transition from a shallow-wrapped to a deep-wrapped state, a very weak increase of $\tilde{w}$ is observed with decreasing $\kappa_{\rm v}/\kappa_{p}$, i.e., the stability of the SW over the DW state increases. Within the shallow-wrapped state, a vesicle shape transformation occurs between soft oblate and stiff prolate vesicles. 

\begin{figure}[htbp!]
    \centering
    \includegraphics[width=\figwidth]{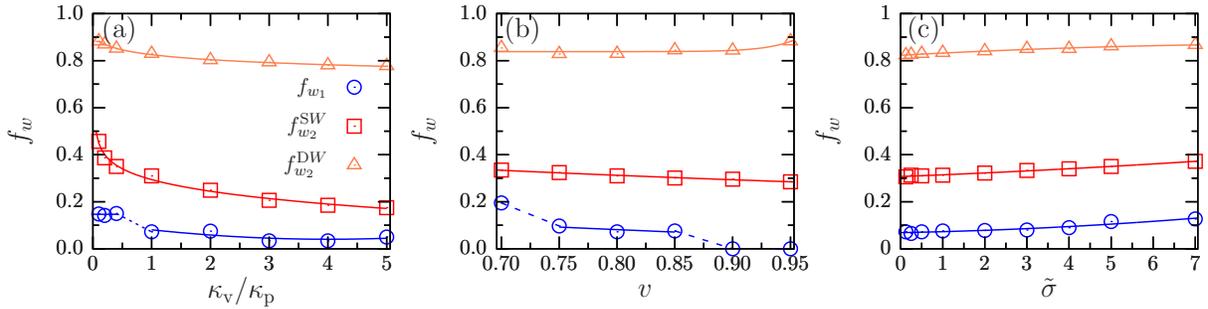}
    \caption{Wrapping fractions $f_w$ for the shallow-wrapped states that coexist with the non-wrapped state at the discontinuous binding transition $W_1$, and for the coexisting shallow-wrapped and deep-wrapped states that coexist at the $W_2$ transition. The wrapping fractions are presented as function of (a) $\kappa_{\rm v}/\kappa_{\rm p}$ for fixed $v=0.8$ and $\tilde{\sigma}=0.5$, (b) $v$ for fixed $\kappa_{\rm v}/\kappa_{\rm p}=1$ and $\tilde{\sigma}=0.5$, and (c) $\tilde{\sigma}$ for fixed $v=0.8$ and $\kappa_{\rm v}/\kappa_{\rm p}=1$. The dashed lines represent discontinuities in $f_{w_1}$ associated with shape changes of the vesicles.}
    \label{fig:wrappingFractions}
\end{figure}

The effect of the tension $\tilde{\sigma}$ of the planar membrane on the wrapping diagram is shown in Fig.~\ref{fig:dEng+phaseDia}(c). The adhesion strength $w_1$ for the binding transition $W_1$ is independent of $\tilde{\sigma}$. The adhesion strength $w_2$ for the shallow- to deep-wrapped transition $W_2$ increases strongly with increasing $\tilde{\sigma}$. The adhesion strength $w_3$ for the envelopment transition $W_3$ increases only weakly for small $\tilde{\sigma}$ and strongly for high $\tilde{\sigma}$. Finally, the reduced volume of the vesicles is varied in the range $0.7 \leq v \leq 0.95$, where free vesicles are prolate, see Fig.~\ref{fig:dEng+phaseDia}(d) and Fig. 1S in the Supporting Information. The adhesion strengths for all three transitions increase with decreasing $v$, accompanied by a change of the shape of the free vesicles that become more elongated with their tips more pointed. Therefore, binding as well as complete wrapping become more difficult, and the stability of the partial-wrapped states increases.

Figure \ref{fig:wrappingFractions} shows the wrapping fractions for partial-wrapped states that coexist with the free vesicles at the $W_1$ transition, and for shallow-wrapped and deep-wrapped states that coexist at the $W_2$ transition. For the $W_2$ transition, the wrapping fractions $f_{w_2}^{\rm SW}$ for the SW states and $f_{w_2}^{\rm DW}$ for the DW state slowly increase with decreasing $\kappa_{\rm v}/\kappa_{\rm p}$ at reduced volume $v=0.8$ and tension $\tilde{\sigma}=0.5$, see Fig.~\ref{fig:wrappingFractions}(a). The stronger increase of $f_{w_2}^{\rm SW}$ indicates the vanishing of DW states in the limit $\kappa_{\rm v}/\kappa_{\rm p} \rightarrow 0$. For the $W_1$ transition, a sudden increase in wrapping fraction of the partial-wrapped states $f_{w_1}$ is observed at $\kappa_{\rm v}/\kappa_{\rm p} \simeq 1$, which is associated with the shape change of the vesicle from prolate to oblate.

The wrapping fractions $f_{w_2}^{\rm SW}$ and $f_{w_2}^{\rm DW}$ remain almost unchanged when the reduced volume is varied in the range $0.7 \leq v \leq 0.95$ at $\kappa_{\rm v}/\kappa_{\rm p}=1$ and $\tilde{\sigma}=0.5$, see Fig.~\ref{fig:wrappingFractions}(b). However, the variation of $f_{w_1}$ with $v$ is discontinuous. A sudden drop of $f_{w_1}$ is observed as $v$ increases from $0.7$ to $0.75$. This is associated with the shape change of the vesicle from oblate to prolate. For $v\geq 0.75$, $f_{w_1}$ decreases monotonically with $v$, and then again a sudden drop is observed from $v=0.85$ to $v=0.9$. Such a behavior is expected as the shape of the free vesicle changes from peanut-like to ellipsoid-like. For $0.75<v<0.85$, the shape of free vesicle is peanut-like (see Fig.~1S), thus, initial attachment to membrane leads to two patches, see Fig.~\ref{fig:snapshots}(a). The vesicle becomes ellipsoid-like for $v \gtrsim 0.9$, thus, initial binding of the vesicle to the planar membrane occurs at the middle part and forms a single patch, see Fig.~2S in the Supporting Information. For both transitions, a weak increase of the wrapping fractions of the coexisting partial-wrapped states is observed with increasing membrane tension $\tilde{\sigma}$ at reduced volume $v=0.8$ and $\kappa_{\rm v}/\kappa_{\rm p}=1$, see Fig.~\ref{fig:wrappingFractions}(c). 

%We find a weak increase of all three wrapping fractions with decreasing $\kappa_{\rm v}/\kappa_{\rm p}$ at reduced volume $v=0.8$ and tension $\tilde{\sigma}=0.5$, see Fig.~\ref{fig:wrappingFractions} (a). For $W_1$ transition, we find a sudden increase in wrapping fraction $f_{w_1}$ at $\kappa_{\rm v}/\kappa_{\rm p}\simeq 1$ which is associated with the shape change of the vesicle from prolate to oblate. The wrapping fraction $f_{\rm w_1}$ of the shallow-wrapped state that coexists with the non-wrapped state at the binding transition $W_1$ increases weakly with decreasing $v$ at $\kappa_{\rm v}/\kappa_{\rm p}=1$ and $\tilde{\sigma}=0.5$, see Fig.~\ref{fig:wrappingFractions}(b). The wrapping fractions of the SW and DW states at the $W_2$ transition remain almost  unchanged when the reduced volume is varied within the range $0.7 \leq v \leq 0.95$ at $\kappa_{\rm v}/\kappa_{\rm p}=1$ and $\tilde{\sigma}=0.5$. For both transitions, a weak increase of $f_w$ with increasing membrane tension $\tilde{\sigma}$ at reduced volume $v=0.8$ and $\kappa_{\rm v}/\kappa_{\rm p}=1$, see Fig.~\ref{fig:wrappingFractions} (c). 

\begin{figure}[htbp!]
    \centering
    \includegraphics[width=\figwidth]{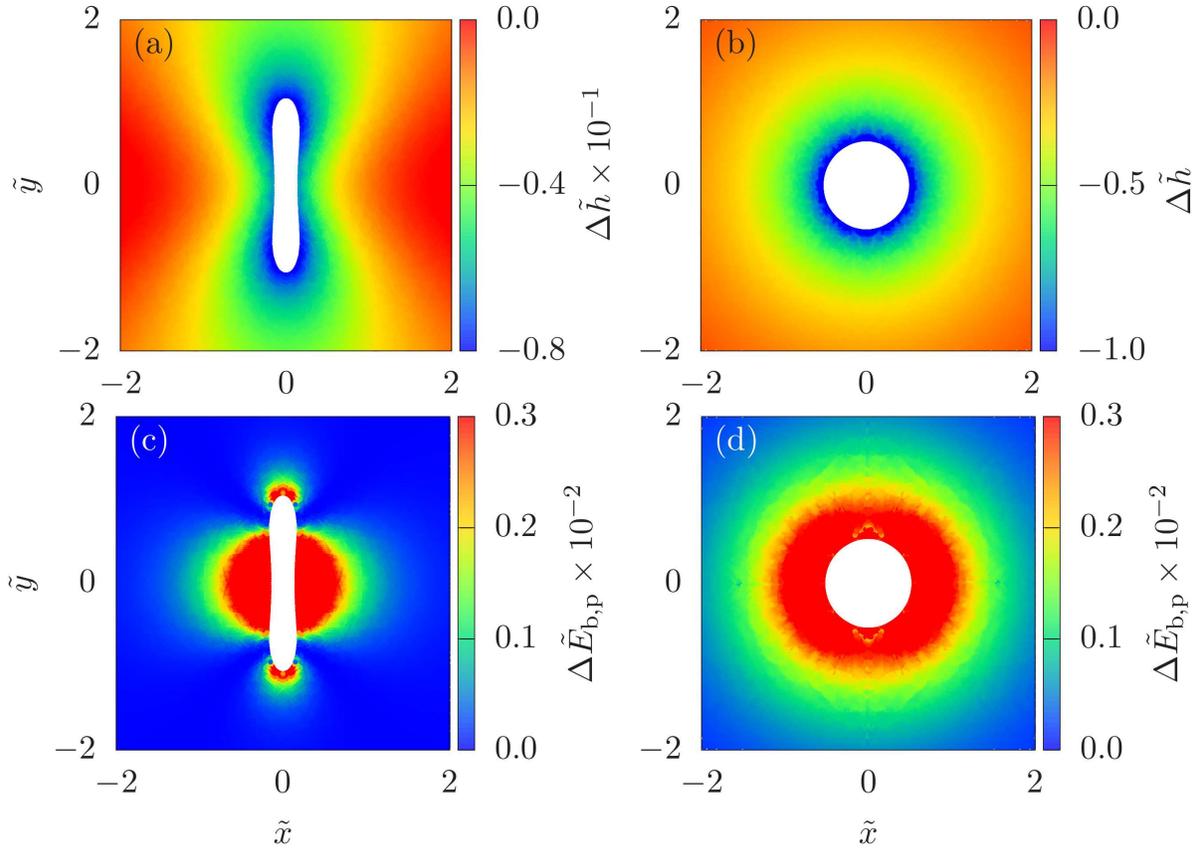}
    \caption{Single vesicles attached from above to initially planar membranes. The heat maps indicate (a,b) the height and (c,d) the bending energy of the free membrane around a vesicle with $v=0.8$, $\kappa_{\rm v}/\kappa_{\rm p}=5$, $\tilde{\sigma}=0.5$ and $\tilde{w}=10.94$ for (a,b) shallow-wrapped and (c,d) deep-wrapped states. The white areas are inside the contact line where the free membrane detaches from the vesicle.}
    \label{fig:dist_h+Eb}
\end{figure}

For all partial-wrapped states, the vesicle deforms the free membrane, see Fig.~\ref{fig:dist_h+Eb}. For partial-wrapped vesicles in submarine orientation, the deviation of the height field describing the free membrane from the plane is strongest around the tips and weakest along the sides. The local bending energy of the free membrane is finite at both, tips and sides, and shows lines of vanishing bending energy separating tips and sides, see Fig.~\ref{fig:dist_h+Eb}(c). The free-membrane deformation at the sides can be approximated by calculations for wrapping infinitely long cylindrical particles \cite{mkrtchyan_adhesion_2010,weikl_indirect_2003}, the  deformation at the tips by calculations for wrapping spherical particles \cite{dasgupta_wrapping_2013,deserno_elastic_2004}. However, the free-membrane deformation along the sides rises the particle height above the optimal value for catenoid formation around the tips.  The finite mean curvature along the sides and at the tips thus originate from principal curvatures with opposite signs. Therefore, the mean curvature and also the local bending energy vanish along lines where the principal curvatures have equal magnitude. For the deep-wrapped vesicles, the free membrane deformation and the bending energy distribution are cylindrical symmetric, see Fig.~\ref{fig:dist_h+Eb}(d). Here, the origin of the finite bending energy of the free membrane is the membrane tension that prevents a catenoidal shape of the neck.

%%%%%%%%%%%%%%%%%%%%%%%%%%%%%%%%%%%%%%%%%%%%%%%%%%%%%%%%%%%%%%%%%%%%%%%%%%%%%%%%%%%%%%%%%%%%%%%%%%%%%%%
\begin{figure}[htbp!]
    \centering
    \includegraphics[width=\figwidth]{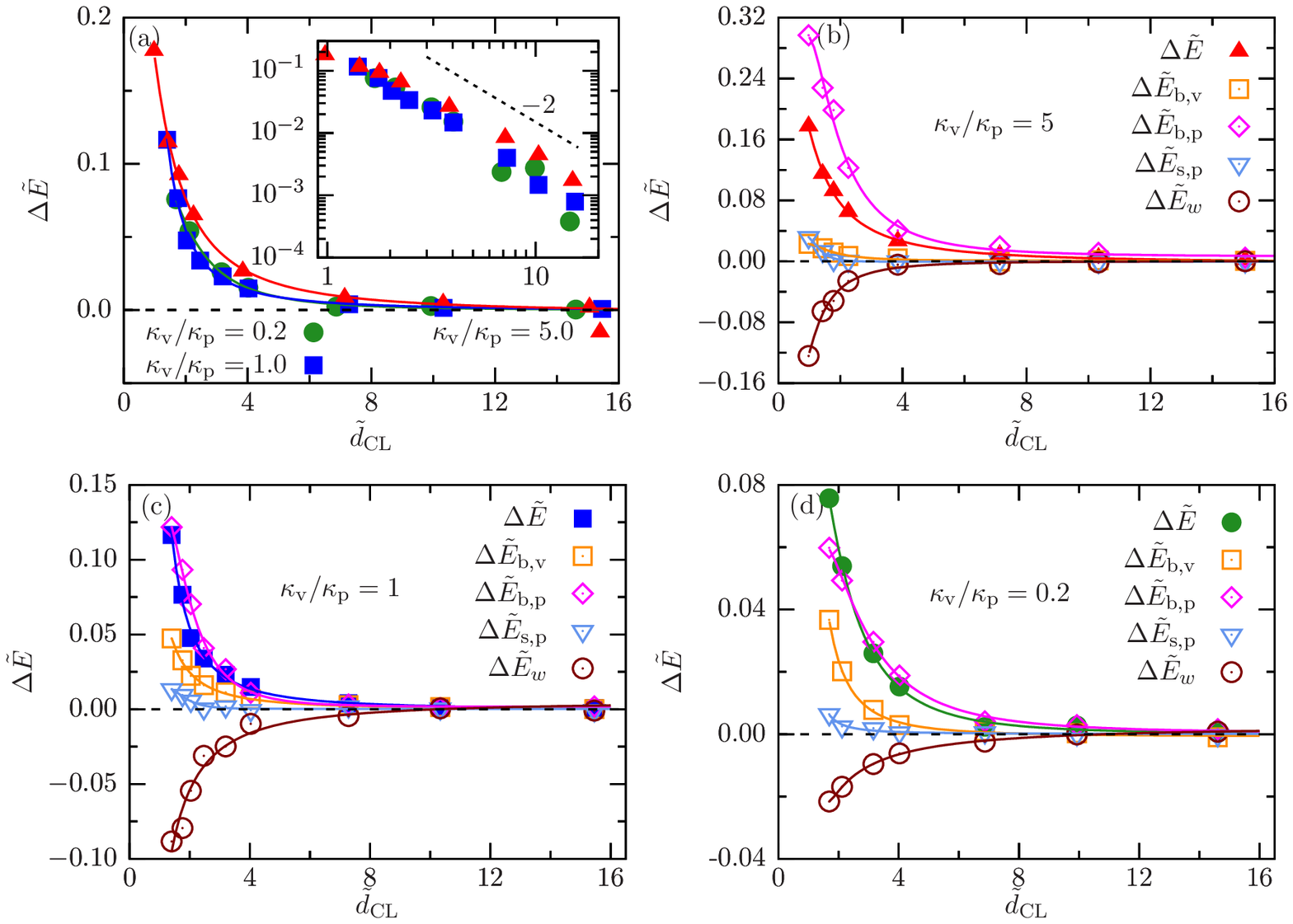}
    \caption{Deep-wrapped vesicles at $\tilde{w}=15$. (a) Interaction potential $\Delta\tilde{E}$ as a function of distance $\tilde{d}_{\rm CL}$ for the bending rigidity ratios $\kappa_{\rm v}/\kappa_{\rm p}=0.2, 1.0$ and $5.0$ at fixed $v=0.8$, and $\tilde{\sigma}=0.5$. Inset shows the same on log-log scale. The solid line presents a power law with an exponent $-2$. The individual energy contributions: the change in bending energy of the vesicle, $\Delta \tilde{E}_{\rm b, v}$, the change in bending energy of the planar membrane, $\Delta \tilde{E}_{\rm b, p}$, the change in surface energy of the planar membrane, $\Delta \tilde{E}_{\rm s,p}$ and the change in adhesion energy, $\Delta\tilde{E}_{\rm w}$, as function of distance $\tilde{d}_{\rm CL}$ are presented for (b) $\kappa_{\rm v}/\kappa_{\rm p}=5.0$ (c) $\kappa_{\rm v}/\kappa_{\rm p}=1.0$ and (d) $\kappa_{\rm v}/\kappa_{\rm p}=0.2$.}
    \label{fig:pten+all-compts-DW}
\end{figure} 
%%%%%%%%%%%%%%%%%%%%%%%%%%%%%%%%%%%%%%%%%%%%%%%%%%%%%%%%%%%%%%%%%%%%%%%%%%%%%%%%%%%%%%%%%%%%%%%%%%%%%%%
{\bf Membrane-mediated interactions between two partial-wrapped vesicles.} The deformation of the initially planar, free membrane induces membrane-mediated interactions between two partial-wrapped vesicles. When the vesicles approach each other, both the initially planar membrane and the attached vesicles deform. We predict membrane-mediated pair interactions between shallow- and deep-wrapped prolate vesicles with $v=0.8$ at initially planar membranes with a small finite membrane tension $\tilde{\sigma}=0.5$. The value of the adhesion strength $\tilde{w}$ is chosen such that either a deep-wrapped or a shallow-wrapped state is stable. The interaction potentials are obtained by calculating the total energies of the system for various distances between the vesicles. The change in the total energy $\Delta \tilde{E}$ is measured with respect to the vesicles at infinite distance, {\it i.e.}, $\Delta \tilde{E}=\tilde{E}(r) - \tilde{E}(\infty)$.   
We calculate the total-energy difference $\Delta \tilde{E}$ as function of the reduced distance $\tilde{d}_{\rm CL}=d_{\rm CL}/a_0$, where $d_{\rm CL}$ is the minimum contactline-to-contactline distance between the two vesicles, and $a_0=\sqrt{A_v/(4 \pi)}$ is the radius of a spherical vesicle with the same membrane area. To understand the origin of the membrane-mediated interaction, we split the total potential energy $\Delta \tilde{E}$ into its individual components, 
\begin{equation}
   \Delta \tilde{E} (\tilde{d}_{\rm CL})=\Delta \tilde{E}_{\rm b,v}(\tilde{d}_{\rm CL})+\Delta \tilde{E}_{\rm b,p}(\tilde{d}_{\rm CL})+\Delta \tilde{E}_{\rm s, p}(\tilde{d}_{\rm CL})+\Delta \tilde{E}_{w}(\tilde{d}_{\rm CL}),
   \label{eq:split_pten}
\end{equation}
where $\Delta \tilde{E}_{\rm b,v}$ is the change of bending energy of the vesicles, $\Delta \tilde{E}_{\rm b,p}$ is the change of bending energy of the planar membrane, $\Delta \tilde{E}_{\rm s, p}$ is the change of tension energy of the planar membrane, and $\Delta \tilde{E}_{w}$ is the change of the adhesion energy. 

For two deep-wrapped vesicles, $\tilde{w} = 15$, the energy $\Delta \tilde{E}$ increases monotonically with decreasing distance between the two vesicles, see Fig.~\ref{fig:pten+all-compts-DW}(a); the vesicles mutually repel each other. We find a purely repulsive interaction with decreasing strength for decreasing bending rigidity of the vesicle membrane. Within the considered distance range, the interaction potential is well described by an effective power law with exponent $-2$, see inset of Fig.~\ref{fig:pten+all-compts-DW}(a). In Fig.~\ref{fig:pten+all-compts-DW}(b)-(d), all energy components of Eq.~(\ref{eq:split_pten}) are plotted as a function of $\tilde{d}_{\rm CL}$ for $\kappa_{\rm v}/\kappa_{\rm p}=0.2$, $1$, and $5$. The bending-energy contributions, $\Delta \tilde{E}_{\rm b,p}$ and $\Delta \tilde{E}_{\rm b,v}$, both increase with decreasing distance between the vesicles. The deformation of the planar membrane increases as the vesicles become stiffer, thus the contribution of $\Delta \tilde{E}_{\rm b,p}$ to the total energy increases with increasing $\kappa_{\rm v}/\kappa_{\rm p}$. Within the range of our parameters, the relative contribution of $\Delta \tilde{E}_{\rm b,v}$ increases as the vesicles become softer, because the deformation of the vesicles becomes less expensive than that of the free membrane. The contribution of the adhesion energy $\Delta \tilde{E}_{w}$ is significant for all cases. We find that $\Delta \tilde{E}_{w}$ becomes more negative at short distances, which is expected as the adhered area increases at short distances; the change of wrapping fraction becomes less important with decreasing $\kappa_{\rm v}/\kappa_{\rm p}$, see Fig.~S12 in the Supporting Information. Thus, the contribution of $\Delta \tilde{E}_{w}$ becomes weaker with decreasing $\kappa_{\rm v}/\kappa_{\rm p}$. For all cases, the repulsive potential mainly originates from $\Delta \tilde{E}_{\rm b,p}$ and $\Delta \tilde{E}_{\rm b,v}$. We observe a weak increase of $\Delta \tilde{E}_{\rm s,p}$ as the distance between the vesicles decreases, which is connected to the increase of adhered membrane area as the vesicles approach each other, see Fig.~S12 in the Supporting Information. The contribution of $\Delta \tilde{E}_{\rm s, p}$ to the total energy is negligible because a very small membrane tension is applied.
%%%%%%%%%%%%%%%%%%%%%%%%%%%%%%%%%%%%%%%%%%%%%%%%%%%%%%%%%%%%%%%%%%%%%%%%%%%%%%%%%%%%%%%%%%%%%%%%%%%%%%%
\begin{figure}[htbp!]
    \centering
    \includegraphics[width=\figwidth]{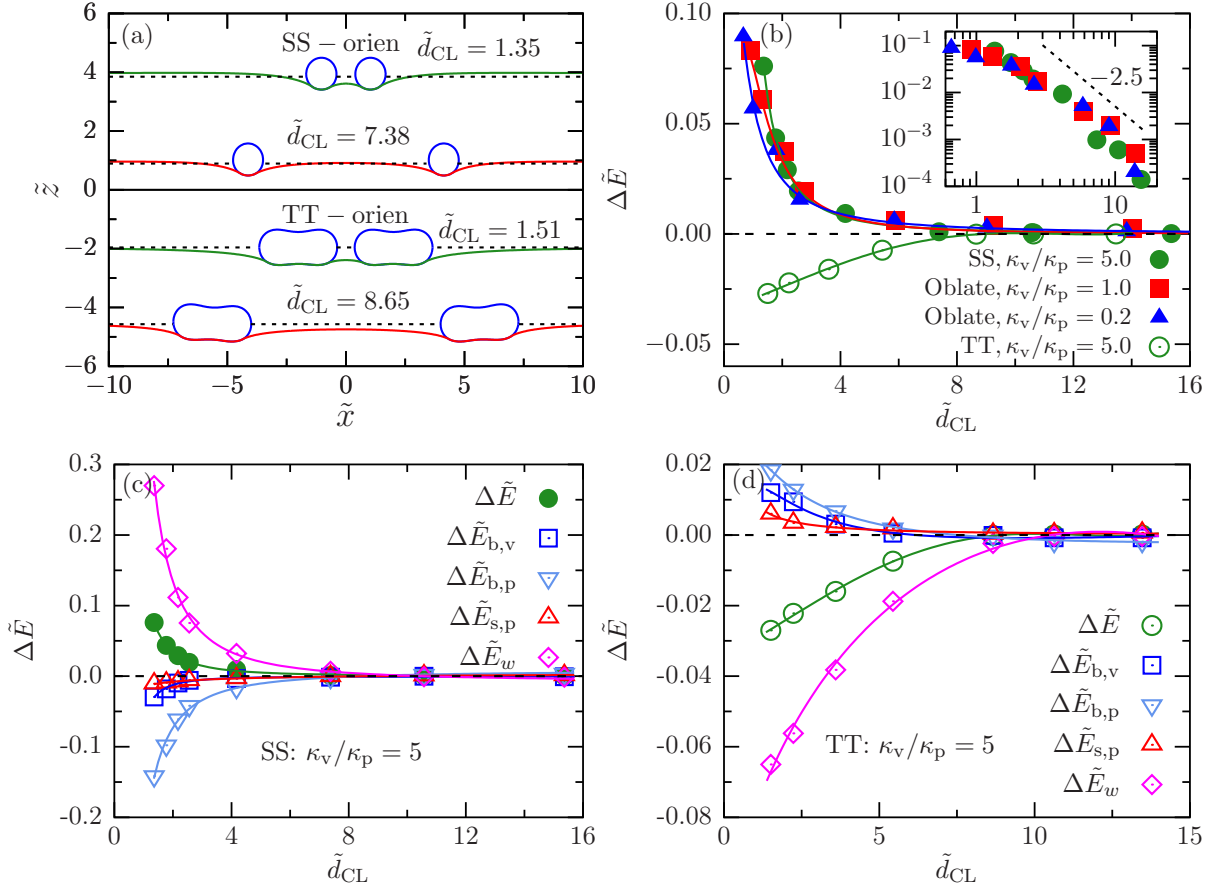}
    \caption{Shallow-wrapped vesicles at $\tilde{w}=10$. (a) Cross-section side view of two vesicles with side-by-side (SS) and tip-to-tip (TT) orientations at  different distances for the choice of parameters $v=0.8$, $\tilde{\sigma=0.5}=0.5$, and $\kappa_{\rm v}/\kappa_{\rm p}=5.0$. (b) Interaction potential $\Delta\tilde{E}$, as a function of the distance $\tilde{d}_{\rm CL}$ between the two vesicles in shallow-wrapped states with SS- and TT-orientations. For SS orientation the interaction potential is plotted for the bending rigidity ratios $\kappa_{\rm v}/\kappa_{\rm p}=0.2, 1.0$ and $5.0$ at $v=0.8$ and $\tilde{\sigma}=0.5$. The inset shows the same on a log-log scale. The solid line presents a power law with an exponent $-2.5$. For TT orientation the interaction potential is shown for $\kappa_{\rm v}/\kappa_{\rm p}=5.0$. The individual energy contributions: the change in bending energy of the vesicle, $\Delta \tilde{E}_{\rm b, v}$, the change in bending energy of the planar membrane, $\Delta \tilde{E}_{\rm b, p}$, the change in surface energy of the planar membrane, $\Delta \tilde{E}_{\rm s,p}$ and the change in adhesion energy, $\Delta\tilde{E}_{\rm w}$, as function of distance $\tilde{d}_{\rm CL}$ are presented at   $\kappa_{\rm v}/\kappa_{\rm p}=5.0$ and $\tilde{\sigma}=0.5$ for (c) SS-orientation and (d) TT-orientation.}
    \label{fig:pten+all-compts-SW}
\end{figure} 
%%%%%%%%%%%%%%%%%%%%%%%%%%%%%%%%%%%%%%%%%%%%%%%%%%%%%%%%%%%%%%%%%%%%%%%%%%%%%%%%%%%%%%%%%%%%%%%%%%%%%%%
%%%%%%%%%%%%%%%%%%%%%%%%%%%%%%%%%%%%%%%%%%%%%%%%%%%%%%%%%%%%%%%

We also investigate the membrane-mediated interaction between two shallow-wrapped vesicles with reduced volume $v=0.8$ and tension $\tilde{\sigma}=0.5$ for the planar membrane, see Fig.~\ref{fig:pten+all-compts-SW}(a). The adhesion strength is fixed at $\tilde{w}=10$, such that the shallow-wrapped state is stable for all considered values of $\kappa_{\rm v}/\kappa_{\rm p}$. For side-by-side (SS) orientation, the interaction is purely repulsive for $\kappa_{\rm v}/\kappa_{\rm p}=0.2, 1$, and $5$, see Fig.~\ref{fig:pten+all-compts-SW}(b). The interaction strength decreases as the vesicles become softer, similar as for deep-wrapped vesicles. A qualitative change in the interaction potential is associated with the shape change of the vesicles from prolate for $\kappa_{\rm v}/\kappa_{\rm p} = 5$ to oblate for $\kappa_{\rm v}/\kappa_{\rm p} \lesssim 1$. The shape change also leads to higher wrapping fractions of the vesicles from $f_w \approx 0.14$ for $\kappa_{\rm v}/\kappa_{\rm b}=5$ to $f_w \approx 0.27$ for $\kappa_{\rm v}/\kappa_{\rm b} = 1$. Similarly, the deformation energy of the planar membrane decreases more as the prolate vesicles approach each other in side-by-side orientation than for the oblate shapes. However, the deformation energy of the free planar membrane, without the adhered part of the planar membrane, increases with increasing $\kappa_{\rm v}/\kappa_{\rm p}$, see Fig.~S11 in the Supporting Information. 
 
For SS orientation and $\kappa_{\rm v}/\kappa_{\rm p}=5$, the high curvature of the free membrane between the particles increases the bending-energy cost for the free membrane, see Fig.~S11 in the Supporting Information. This leads to partial unwrapping of the vesicles from the planar membrane, see Fig.~\ref{fig:pten+all-compts-SW}(c) and Fig.~S12 in the Supplementary Information. We find both 
bending-energy contributions, $\Delta \tilde{E}_{\rm b,p}$ and $\Delta \tilde{E}_{\rm b,v}$, to become more negative with decreasing distance. The change of adhesion energy $\Delta \tilde{E}_{w}$ is increasing with decreasing vesicle-vesicle distance, which is the dominating contribution causing the repulsion between the vesicles. The contribution of $\Delta {\tilde E}_{\rm s,p}$ to the total energy decreases with decreasing $\tilde{d}_{\rm CL}$, which is expected because the total membrane area decreases as two vesicles approach each other. However, the contribution of  $\Delta {\tilde E}_{\rm s,p}$ to the total energy is negligible. The total energy as function of the distance can be described by an effective power law $\tilde{d}_{\rm CL}^{-2.5}$, see the inset of Fig.~\ref{fig:pten+all-compts-DW}(b). Thus, a smaller area of the vesicles is adhered to the planar membrane as they approach each other, opposite to deep-wrapped vesicles where the wrapping fraction increases. As a result, a qualitative change is observed in the behavior of the individual components in comparison with deep-wrapped vesicles. Our shallow-wrapped vesicles in SS orientation experience stiffer membrane-mediated interaction potentials than our deep-wrapped vesicles.

\begin{figure}[htbp!]
    \centering
    \includegraphics[width=\figwidth]{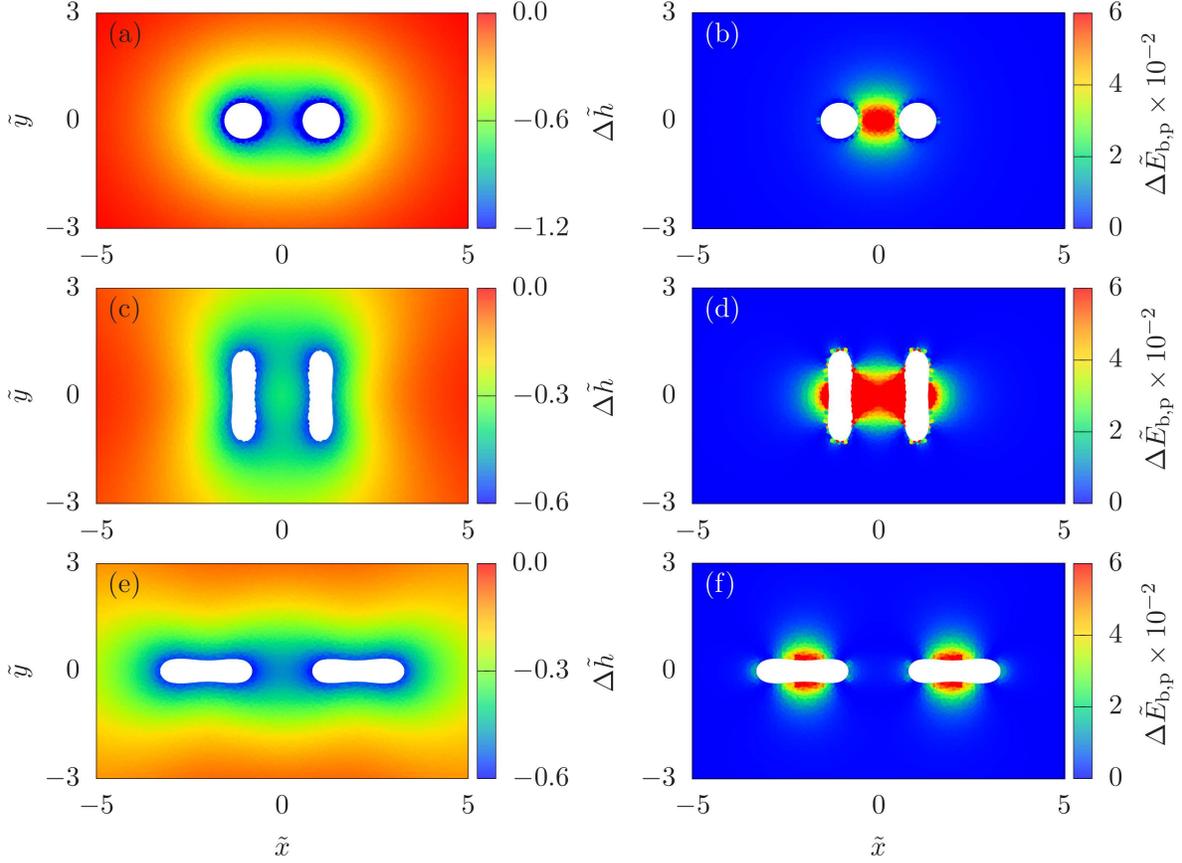}
    \caption{Free-membrane deformation for two vesicles at a planar membrane. (a,c,e) Local membrane height difference $\Delta \tilde{h} = [h-h_0]/a_0$ with the local membrane height $h$ and the height $h_0$ of the wire frame spanning the membrane patch, and (b,d,f) local bending energy of the free membrane. The values are shown in the $x-y$ plane for vesicles with $v=0.8$, reduced membrane tension $\tilde{\sigma}=0.5$, and $\kappa_{\rm v}/\kappa_{\rm p}=5.0$. (a, b) Deep-wrapped vesicles with their centers of mass at $(x_1,y_1)=(-x_2,y_2)=(0.49,0)$, (c,d) shallow-wrapped in side-by-side orientation with their centers of mass at $(x_1,y_1)=(-x_2,y_2)=(0.68,0)$, and (e, f) shallow-wrapped in tip-to-tip orientation with their centers of mass at $(x_1,y_1)=(-x_2,y_2)=(0.75,0)$.}
    \label{fig:dist-h+Eb-DW+SW-SS+TT}
\end{figure}

For tip-to-tip (TT) orientation and $\kappa_{\rm v}/\kappa_{\rm p}=5$, a cooperative deformation of the membrane by the two vesicles leads to the formation of a joint "trough" for both particles without a strong increase of the deformation-energy cost of the free membrane, see Fig.~\ref{fig:pten+all-compts-SW}(a) and Fig.~\ref{fig:dist-h+Eb-DW+SW-SS+TT}(e,f). The interaction is weakly attractive. To understand this behavior, we again look into the individual energy components, see Fig.~\ref{fig:pten+all-compts-SW}(d). In contrast to the SS orientation, here we find that the wrapping fraction of the vesicles increases as the vesicles approach each other. As a result, tension- and 
bending-energy contributions for both the planar membrane and the vesicles increase as the distance between the vesicles decreases. The effective attraction originates from the gain in adhesion energy. For shallow-wrapped states and $\kappa_{\rm v}/\kappa_{\rm p}\lesssim 1$ the vesicle shape changes from prolate to oblate; therefore, there is no distinction between TT and SS orientations. 

Now, we discuss shapes and local bending energies of the free membrane for both, deep-wrapped and shallow-wrapped vesicles, see Fig.~\ref{fig:dist-h+Eb-DW+SW-SS+TT}(a)-(f). At the contactline, where the free membrane detaches from the vesicle, the membrane height deviates most from the height of the wire frame. For two deep-wrapped vesicles with contactline-contactline distance $\tilde{d}_{\rm CL}=0.97$, the free-membrane height also strongly deviates from the wire frame height in the middle between the vesicles, see Fig.~\ref{fig:dist-h+Eb-DW+SW-SS+TT}(a). Here, the deformation is strong because the membrane aims to decrease the high local bending-energy cost, see Fig.~\ref{fig:dist-h+Eb-DW+SW-SS+TT}(b); yet, this energy cost causes the repulsive interaction between the vesicles, compare Fig.~\ref{fig:pten+all-compts-DW}(b). The repulsion leads to tilting of the vesicles, see Fig.~S7 in the Supplementary Information.  
Interestingly, the membrane-mediated interaction induces an increase of the center-of-mass height of the vesicles with respect to the wire frame, see Fig.~S8 in the Supplementary Information.

For two shallow-wrapped vesicles with contactline-contactline distance $\tilde{d}_{\rm CL}=1.36$, the height of the free membrane deviates most strongly from the wire frame height at the tips of the vesicles, see Fig.~\ref{fig:dist-h+Eb-DW+SW-SS+TT}(c). The bending energy cost between the vesicles is very high for SS orientation, see Fig.~\ref{fig:dist-h+Eb-DW+SW-SS+TT}(d), which leads to a partial unwrapping of the vesicle sides that face each other and therefore a repulsive interaction because of the decreasing adhesion energy when the vesicles approach each other, see Fig.~\ref{fig:pten+all-compts-SW}(c) and Fig.~S12 in the Supplementary Information. For TT orientation, the bending energy of the membrane between the vesicles is very small although the membrane is also strongly deformed, see  Fig.~\ref{fig:dist-h+Eb-DW+SW-SS+TT}(e). The membrane height deviates strongest from the wire-frame height at the tips that face each other--and wrapping both vesicles together increases the wrapping at the sides as well. However, the overall increase of the bending energy in TT orientation is much lower than in SS orientation; the joint deformation of the membrane by both vesicles strongly increases the wrapping at their sides and thereby the adhesion energy $E_{w}$, see  Fig.~\ref{fig:dist-h+Eb-DW+SW-SS+TT}(f) and Fig.~\ref{fig:pten+all-compts-SW}(d).
Also for shallow-wrapped vesicles, the heights of the centers of mass of the vesicles decrease with decreasing distance, see Fig.~S8(b) in the Supporting Information. 

%%%%%%%%%%%%%%%%%%%%%%%%%%%%%%%%%%%%%%%%%%%%%%%%%%%%%%%%%%%%%%%
 
%Deformation and adhesion energies, as well as vesicle shapes, have been predicted for adhesion of vesicles to various substrates--from planar and nanostructured surfaces to nanoparticles. 
%Osmotic pressure plays a particularly important role for high osmotic concentrations under physiological conditions, where also non-spherical vesicles can be fabricated. Whereas for high osmotic concentrations initially spherical vesicles may mimic a hard particles, small osmotic concentration can be used to control the vesicle deformability. Osmotic concentrations can be used to control vesicle volume and shape. Adjusting the reduced  volume can be used to obtain a variety of vesicle shapes, in particular prolates and oblates.

\section*{Conclusions}
In this work, we have predicted the wrapping behavior and calculated membrane-mediated interactions between prolate vesicles at planar membranes. For single-vesicle systems, we systematically varied the reduced volume $v$ of the vesicles, the bending rigidity ratio $\kappa_{\rm v}/\kappa_{\rm p}$, and the  membrane tension $\tilde{\sigma}$ of the planar membrane. Wrapping always starts from the minimum curvature regions of the vesicles. Thus, initial wrapping for low adhesion strengths occurs in submarine orientation where the major axis of the vesicle is parallel to the planar membrane. With increasing adhesion strength, for stiff vesicles the orientation changes from submarine to rocket, where the major axis of the vesicle is perpendicular to the planar membrane. For soft vesicles, a shape and orientation change of partial-wrapped vesicles from prolate submarine to oblate to prolate rocket can be observed. The binding transition, the shallow-to-deep-wrapped transition, and the envelopment transition are located and characterized by systematically varying the adhesion strength. We have shown that the binding and the shallow-to-deep-wrapped transition are always discontinuous, and the envelopment transition is always continuous. For fixed $v$ and $\tilde{\sigma}$, we have shown that the binding transition shifts to lower adhesion strengths and the envelopment transition shifts to higher adhesion strengths with decreasing bending-rigidity ratio $\kappa_{\rm v}/\kappa_{\rm p}$. The softer the vesicle, the easier it attaches to the membrane, but the more difficult it gets completely wrapped--which is a generic behavior of soft, deformable particles. For fixed $v$ and $\kappa_{\rm v}/\kappa_{\rm p}$, we have shown that the binding transition is independent of the membrane tension. An increase of membrane tension stabilizes shallow-wrapped states over deep-wrapped and complete-wrapped states. 

The membrane-mediated interaction between two shallow-wrapped or two deep-wrapped prolate vesicles ($v=0.8$) at planar membranes is studied by systematically varying the distance between two partial-wrapped vesicles. Particle systems are more complex than spherical-cap inclusions\cite{reynwar_membrane-mediated_2011,reynwar_aggregation_2007,kim_curvature-mediated_1998,weikl_interaction_1998,goulian_long-range_1993}, curved membrane domains \cite{baumgart2003imaging}, and curved proteins\cite{noguchi_membrane_2017} for which the wrapping fraction does not depend on the distance. For deep-wrapped states of our prolate vesicles, we found the interaction potential to be repulsive, which originates from the deformation of both, free membrane and vesicles. The strength of the repulsive interaction decreases with the decreasing $\kappa_{\rm v}/\kappa_{\rm p}$, {\it i.e.}, with increasing softness of the vesicles. For shallow-wrapped states, a qualitative change in the interaction potential from attractive in tip-to-tip orientation to repulsive in side-by-side orientation is observed. For side-by-side orientation, the interaction potential is purely repulsive and becomes weaker as the vesicles become softer, which is associated with the shape change of the vesicles from prolate to oblate. Although, the interaction for both, deep-wrapped vesicles and shallow-wrapped vesicles in SS orientation, is repulsive, the reason is different in both cases. For side-by-side orientation, the repulsive interaction originates from the loss of adhesion energy as the two vesicles approach each other. For tip-to-tip orientation, the interaction energy is attractive, which originates from the gain in adhesion energy as the vesicles approach each other.  

Our theoretical predictions for wrapping and for pair-interactions of soft particles at membranes can be used to optimise the shapes and the elastic properties of deformable particles for efficient use in nanomedicine, such as for targeted drug delivery. It will be also interesting to investigate membrane-mediated interactions between many particles, which will help us to understand more complex problems like aggregation of virions on lipid-bilayer membranes. The understanding of the wrapping behavior of the vesicle-membrane systems is leading off to investigate the interaction of polymeric particles, {\it e.g.}, star polymers, polymer-grafted nanoparticles, and microgels, with lipid bilayer membranes. However, contrary to vesicles polymeric particles have bulk and shear elasticities, and their surface area increases upon binding to a membrane \cite{scotti2206}. Yet the knowledge that we gained from non-spherical vesicle-membrane systems provides a starting point to estimate the membrane-mediated interaction for anisotropic soft particles in general.

\section{Model and methods}
\label{sec:methods} 
The continuum model in Eq.~(\ref{Eq:Helfrich_Hamiltonian}) is limited to systems with particle sizes that are at least few times larger than the thickness of a lipid-bilayer membranes; bending and tension energy both contribute in the wrapping process. The total energy of the system can be expressed as a dimensionless quantity by dividing by the wrapping energy of spherical particles, $\pi\kappa_{\rm p}$, and expressing all length scales in terms of the membrane area $A_{\rm v}$ of the vesicle. This gives
\begin{equation}
\tilde{E}=\frac{E}{\pi\kappa_{\rm p}}=\frac{2}{\pi A_{\rm v}} \left[ \int_{A_{\rm p}} dS [A_{\rm v} H^2 + \pi \tilde{\sigma}] + \frac {\kappa_{\rm v}}{\kappa_{\rm p}} \int_{A_{\rm v}} dS  A_{\rm v}H^2 \right] - \tilde{w} \int_{A_{\rm ad}} \frac {dS}{A_{\rm v}} \, ,
\label{Eq:scaled_Helfrich_Hamiltonian}
\end{equation}
where $\tilde{\sigma} = \sigma A_{\rm v}/(2\pi\kappa_{\rm p})$ and $\tilde{w}=wA_{\rm v}/(\pi\kappa_{\rm p})$ are the reduced surface tension and reduced adhesion strength, respectively. 

The integrals are discretized using triangulated membranes \cite{gompper_triangulated-surface_2004,sunil_kumar_budding_2001,kroll_conformation_1992}. For energy minimization, we employ the freely available program package “Surface Evolver”, where a discretized surface is formed of vertices, edges, and facets \cite{brakke_surface_1992}. To calculate minimal-energy shapes and associated deformation energies, we refine the triangulation and minimize the energy at each refinement level in an alternating sequence. The refinement of the triangles and energy minimization continue until the desired accuracy is achieved. The enclosed volumes ($V_{\rm v}$) and vesicle areas ($A_{\rm v}$) are fixed with the help of Lagrange multipliers. The vesicle shapes are controlled via the reduced volume $v$. The upper bound $v=1$ corresponds to a spherical vesicle, while $v<1$ represents nonspherical vesicles, see Fig.~1S in the Supporting Information.  

The numerical data for the deformation energy are fitted using a piecewise function, independently for submarine and rocket orientation. In each region, we fit the energy using a fourth-order polynomial
\begin{equation}
    f(f_w) = \sum_{i=0}^{4} a_0 [f_{w}]^i
    \label{Eq:4thPolynomial}
\end{equation}
with the coefficients $a_{0},..., a_{4}$. For the calculation of the phase diagrams we use the sum of deformation and adhesion energy, where the latter is proportional to the wrapping fraction. To locate different transitions, {\it i.e.} binding transitions, shallow-wrapped to deep-wrapped transitions, and envelopment transitions, we take the first derivative of the deformation energy with respect to the wrapping fraction for different adhesion strengths. The zeros of the first derivative indicate the local minima and maxima of the wrapping energies.

To calculate membrane-mediated pair interaction potentials, two vesicles are placed on the planar membrane at fixed center-of-mass distances $\tilde{d}_{\rm CC}=d_{\rm CC}/a_0$, where $a_0=\sqrt{(A_{\rm v}/4\pi)}$. However, for the presentation of the results, we use the contactline-to-contactline distance $\tilde{d}_{\rm CL}=d_{\rm CL}/a_0$; we have also calculated the vesicle surface-to-surface distances $\tilde{d}_{\rm SS}=d_{\rm SS}/a_0$, see Table~S2 in the Supporting Information. For each distance, the total energy is minimized and the deformation energy $\Delta \tilde{E}$ as function of the wrapping fraction $f_w$ is obtained. We exploit the mirror symmetry and calculate only a quarter or half of the actual system. For the deep-wrapped states, the contactline is stabilized by forcing it to a plane with an arbitrary tilt angle. As for the single-vesicle systems, the  deformation energy is fit using fourth-order polynomials. The adhesion strength $\tilde{w}$ is chosen such that the minimum of $\Delta \tilde{E}$ is either a stable deep-wrapped or a stable shallow-wrapped state for all vesicle-vesicle distances. 
 
\begin{acknowledgement}

We gratefully acknowledge financial support from the Deutsche Forschungsgemeinschaft (DFG) within the SFB 985 “Functional Microgels and Microgel Systems” and helpful discussions with Jerome Crassous (Aachen) and Ken Brakke (Selinsgrove).

\end{acknowledgement}

\bibliography{JM-TA-GG}

\providecommand{\latin}[1]{#1}
\makeatletter
\providecommand{\doi}
  {\begingroup\let\do\@makeother\dospecials
  \catcode`\{=1 \catcode`\}=2 \doi@aux}
\providecommand{\doi@aux}[1]{\endgroup\texttt{#1}}
\makeatother
\providecommand*\mcitethebibliography{\thebibliography}
\csname @ifundefined\endcsname{endmcitethebibliography}
  {\let\endmcitethebibliography\endthebibliography}{}
\begin{mcitethebibliography}{53}
\providecommand*\natexlab[1]{#1}
\providecommand*\mciteSetBstSublistMode[1]{}
\providecommand*\mciteSetBstMaxWidthForm[2]{}
\providecommand*\mciteBstWouldAddEndPuncttrue
  {\def\EndOfBibitem{\unskip.}}
\providecommand*\mciteBstWouldAddEndPunctfalse
  {\let\EndOfBibitem\relax}
\providecommand*\mciteSetBstMidEndSepPunct[3]{}
\providecommand*\mciteSetBstSublistLabelBeginEnd[3]{}
\providecommand*\EndOfBibitem{}
\mciteSetBstSublistMode{f}
\mciteSetBstMaxWidthForm{subitem}{(\alph{mcitesubitemcount})}
\mciteSetBstSublistLabelBeginEnd
  {\mcitemaxwidthsubitemform\space}
  {\relax}
  {\relax}

\bibitem[Dasgupta \latin{et~al.}(2014)Dasgupta, Auth, Gov, Satchwell, Hanssen,
  Zuccala, Riglar, Toye, Betz, Baum, and
  Gompper]{dasgupta_membrane-wrapping_2014}
Dasgupta,~S.; Auth,~T.; Gov,~N.; Satchwell,~T.; Hanssen,~E.; Zuccala,~E.;
  Riglar,~D.; Toye,~A.; Betz,~T.; Baum,~J.; Gompper,~G. Membrane-{Wrapping}
  {Contributions} to {Malaria} {Parasite} {Invasion} of the {Human}
  {Erythrocyte}. \emph{Biophys. J.} \textbf{2014}, \emph{107}, 43--54\relax
\mciteBstWouldAddEndPuncttrue
\mciteSetBstMidEndSepPunct{\mcitedefaultmidpunct}
{\mcitedefaultendpunct}{\mcitedefaultseppunct}\relax
\EndOfBibitem
\bibitem[Dasgupta \latin{et~al.}(2017)Dasgupta, Auth, and
  Gompper]{dasgupta_nano-_2017}
Dasgupta,~S.; Auth,~T.; Gompper,~G. Nano- and microparticles at fluid and
  biological interfaces. \emph{J. Phys.: Condens. Matter} \textbf{2017},
  \emph{29}, 373003\relax
\mciteBstWouldAddEndPuncttrue
\mciteSetBstMidEndSepPunct{\mcitedefaultmidpunct}
{\mcitedefaultendpunct}{\mcitedefaultseppunct}\relax
\EndOfBibitem
\bibitem[Joshi \latin{et~al.}(2020)Joshi, de~Beer, Giepmans, and
  Zuhorn]{joshi_endocytosis_2020}
Joshi,~B.~S.; de~Beer,~M.~A.; Giepmans,~B. N.~G.; Zuhorn,~I.~S. Endocytosis of
  {Extracellular} {Vesicles} and {Release} of {Their} {Cargo} from {Endosomes}.
  \emph{ACS Nano} \textbf{2020}, \emph{14}, 4444--4455\relax
\mciteBstWouldAddEndPuncttrue
\mciteSetBstMidEndSepPunct{\mcitedefaultmidpunct}
{\mcitedefaultendpunct}{\mcitedefaultseppunct}\relax
\EndOfBibitem
\bibitem[Sebastiani \latin{et~al.}(2021)Sebastiani, Yanez~Arteta, Lerche,
  Porcar, Lang, Bragg, Elmore, Krishnamurthy, Russell, Darwish, Pichler,
  Waldie, Moulin, Haertlein, Forsyth, Lindfors, and
  Cárdenas]{sebastiani_apolipoprotein_2021}
Sebastiani,~F. \latin{et~al.}  Apolipoprotein {E} {Binding} {Drives}
  {Structural} and {Compositional} {Rearrangement} of {mRNA}-{Containing}
  {Lipid} {Nanoparticles}. \emph{ACS Nano} \textbf{2021}, \emph{15},
  6709--6722\relax
\mciteBstWouldAddEndPuncttrue
\mciteSetBstMidEndSepPunct{\mcitedefaultmidpunct}
{\mcitedefaultendpunct}{\mcitedefaultseppunct}\relax
\EndOfBibitem
\bibitem[Pardi \latin{et~al.}(2015)Pardi, Tuyishime, Muramatsu, Kariko, Mui,
  Tam, Madden, Hope, and Weissman]{pardi_expression_2015}
Pardi,~N.; Tuyishime,~S.; Muramatsu,~H.; Kariko,~K.; Mui,~B.~L.; Tam,~Y.~K.;
  Madden,~T.~D.; Hope,~M.~J.; Weissman,~D. Expression kinetics of
  nucleoside-modified {mRNA} delivered in lipid nanoparticles to mice by
  various routes. \emph{J. Controlled Release} \textbf{2015}, \emph{217},
  345--351\relax
\mciteBstWouldAddEndPuncttrue
\mciteSetBstMidEndSepPunct{\mcitedefaultmidpunct}
{\mcitedefaultendpunct}{\mcitedefaultseppunct}\relax
\EndOfBibitem
\bibitem[Maier-Hauff \latin{et~al.}(2011)Maier-Hauff, Ulrich, Nestler, Niehoff,
  Wust, Thiesen, Orawa, Budach, and Jordan]{maier-hauff_efficacy_2011}
Maier-Hauff,~K.; Ulrich,~F.; Nestler,~D.; Niehoff,~H.; Wust,~P.; Thiesen,~B.;
  Orawa,~H.; Budach,~V.; Jordan,~A. Efficacy and safety of intratumoral
  thermotherapy using magnetic iron-oxide nanoparticles combined with external
  beam radiotherapy on patients with recurrent glioblastoma multiforme. \emph{J
  Neurooncol} \textbf{2011}, \emph{103}, 317--324\relax
\mciteBstWouldAddEndPuncttrue
\mciteSetBstMidEndSepPunct{\mcitedefaultmidpunct}
{\mcitedefaultendpunct}{\mcitedefaultseppunct}\relax
\EndOfBibitem
\bibitem[Dasgupta \latin{et~al.}(2014)Dasgupta, Auth, and
  Gompper]{dasgupta_shape_2014}
Dasgupta,~S.; Auth,~T.; Gompper,~G. Shape and {Orientation} {Matter} for the
  {Cellular} {Uptake} of {Nonspherical} {Particles}. \emph{Nano Lett.}
  \textbf{2014}, \emph{14}, 687--693\relax
\mciteBstWouldAddEndPuncttrue
\mciteSetBstMidEndSepPunct{\mcitedefaultmidpunct}
{\mcitedefaultendpunct}{\mcitedefaultseppunct}\relax
\EndOfBibitem
\bibitem[Yi \latin{et~al.}(2011)Yi, Shi, and Gao]{yi_cellular_2011}
Yi,~X.; Shi,~X.; Gao,~H. Cellular {Uptake} of {Elastic} {Nanoparticles}.
  \emph{Phys. Rev. Lett.} \textbf{2011}, \emph{107}, 098101\relax
\mciteBstWouldAddEndPuncttrue
\mciteSetBstMidEndSepPunct{\mcitedefaultmidpunct}
{\mcitedefaultendpunct}{\mcitedefaultseppunct}\relax
\EndOfBibitem
\bibitem[Smith and Seifert(2005)Smith, and Seifert]{smith_effective_2005}
Smith,~A.-S.; Seifert,~U. Effective adhesion strength of specifically bound
  vesicles. \emph{Phys. Rev. E} \textbf{2005}, \emph{71}, 061902\relax
\mciteBstWouldAddEndPuncttrue
\mciteSetBstMidEndSepPunct{\mcitedefaultmidpunct}
{\mcitedefaultendpunct}{\mcitedefaultseppunct}\relax
\EndOfBibitem
\bibitem[Dasgupta \latin{et~al.}(2013)Dasgupta, Auth, and
  Gompper]{dasgupta_wrapping_2013}
Dasgupta,~S.; Auth,~T.; Gompper,~G. Wrapping of ellipsoidal nano-particles by
  fluid membranes. \emph{Soft Matter} \textbf{2013}, \emph{9}, 5473--5482\relax
\mciteBstWouldAddEndPuncttrue
\mciteSetBstMidEndSepPunct{\mcitedefaultmidpunct}
{\mcitedefaultendpunct}{\mcitedefaultseppunct}\relax
\EndOfBibitem
\bibitem[Welsch \latin{et~al.}(2010)Welsch, Kolesnikova, Krähling, Riches,
  Becker, and Briggs]{welsch_electron_2010}
Welsch,~S.; Kolesnikova,~L.; Krähling,~V.; Riches,~J.~D.; Becker,~S.;
  Briggs,~J. A.~G. Electron {Tomography} {Reveals} the {Steps} in {Filovirus}
  {Budding}. \emph{PLoS Pathog.} \textbf{2010}, \emph{6}, e1000875\relax
\mciteBstWouldAddEndPuncttrue
\mciteSetBstMidEndSepPunct{\mcitedefaultmidpunct}
{\mcitedefaultendpunct}{\mcitedefaultseppunct}\relax
\EndOfBibitem
\bibitem[Del~Rosario \latin{et~al.}(2019)Del~Rosario, Periz, Pavlou, Lyth,
  Latorre‐Barragan, Das, Pall, Stortz, Lemgruber, Whitelaw, Baum, Tardieux,
  and Meissner]{del_rosario_apicomplexan_2019}
Del~Rosario,~M.; Periz,~J.; Pavlou,~G.; Lyth,~O.; Latorre‐Barragan,~F.;
  Das,~S.; Pall,~G.~S.; Stortz,~J.~F.; Lemgruber,~L.; Whitelaw,~J.~A.;
  Baum,~J.; Tardieux,~I.; Meissner,~M. Apicomplexan {F}‐actin is required for
  efficient nuclear entry during host cell invasion. \emph{EMBO reports}
  \textbf{2019}, \emph{20}\relax
\mciteBstWouldAddEndPuncttrue
\mciteSetBstMidEndSepPunct{\mcitedefaultmidpunct}
{\mcitedefaultendpunct}{\mcitedefaultseppunct}\relax
\EndOfBibitem
\bibitem[Klein \latin{et~al.}(2020)Klein, Cortese, Winter, Wachsmuth-Melm,
  Neufeldt, Cerikan, Stanifer, Boulant, Bartenschlager, and
  Chlanda]{klein_sars-cov-2_2020}
Klein,~S.; Cortese,~M.; Winter,~S.~L.; Wachsmuth-Melm,~M.; Neufeldt,~C.~J.;
  Cerikan,~B.; Stanifer,~M.~L.; Boulant,~S.; Bartenschlager,~R.; Chlanda,~P.
  {SARS}-{CoV}-2 structure and replication characterized by in situ
  cryo-electron tomography. \emph{Nat. Commun.} \textbf{2020}, \emph{11},
  1--10\relax
\mciteBstWouldAddEndPuncttrue
\mciteSetBstMidEndSepPunct{\mcitedefaultmidpunct}
{\mcitedefaultendpunct}{\mcitedefaultseppunct}\relax
\EndOfBibitem
\bibitem[Wang \latin{et~al.}(2019)Wang, Mihut, Rieloff, Dabkowska, Månsson,
  Immink, Sparr, and Crassous]{wang_assembling_2019}
Wang,~M.; Mihut,~A.~M.; Rieloff,~E.; Dabkowska,~A.~P.; Månsson,~L.~K.;
  Immink,~J.~N.; Sparr,~E.; Crassous,~J.~J. Assembling responsive microgels at
  responsive lipid membranes. \emph{Proc. Natl. Acad. Sci. U.S.A.}
  \textbf{2019}, \emph{116}, 5442--5450\relax
\mciteBstWouldAddEndPuncttrue
\mciteSetBstMidEndSepPunct{\mcitedefaultmidpunct}
{\mcitedefaultendpunct}{\mcitedefaultseppunct}\relax
\EndOfBibitem
\bibitem[Likos \latin{et~al.}(1998)Likos, Löwen, Watzlawek, Abbas,
  Jucknischke, Allgaier, and Richter]{likos_star_1998}
Likos,~C.~N.; Löwen,~H.; Watzlawek,~M.; Abbas,~B.; Jucknischke,~O.;
  Allgaier,~J.; Richter,~D. Star {Polymers} {Viewed} as {Ultrasoft} {Colloidal}
  {Particles}. \emph{Phys. Rev. Lett.} \textbf{1998}, \emph{80},
  4450--4453\relax
\mciteBstWouldAddEndPuncttrue
\mciteSetBstMidEndSepPunct{\mcitedefaultmidpunct}
{\mcitedefaultendpunct}{\mcitedefaultseppunct}\relax
\EndOfBibitem
\bibitem[Midya \latin{et~al.}(2020)Midya, Rubinstein, Kumar, and
  Nikoubashman]{midya_structure_2020}
Midya,~J.; Rubinstein,~M.; Kumar,~S.~K.; Nikoubashman,~A. Structure of
  {Polymer}-{Grafted} {Nanoparticle} {Melts}. \emph{ACS Nano} \textbf{2020},
  \emph{14}, 15505--15516\relax
\mciteBstWouldAddEndPuncttrue
\mciteSetBstMidEndSepPunct{\mcitedefaultmidpunct}
{\mcitedefaultendpunct}{\mcitedefaultseppunct}\relax
\EndOfBibitem
\bibitem[Donath \latin{et~al.}(1998)Donath, Sukhorukov, Caruso, Davis, and
  Möhwald]{donath_novel_1998}
Donath,~E.; Sukhorukov,~G.~B.; Caruso,~F.; Davis,~S.~A.; Möhwald,~H. Novel
  {Hollow} {Polymer} {Shells} by {Colloid}-{Templated} {Assembly} of
  {Polyelectrolytes}. \emph{Angew. Chem. Int. Ed.} \textbf{1998}, \emph{37},
  2201--2205\relax
\mciteBstWouldAddEndPuncttrue
\mciteSetBstMidEndSepPunct{\mcitedefaultmidpunct}
{\mcitedefaultendpunct}{\mcitedefaultseppunct}\relax
\EndOfBibitem
\bibitem[Tomalia \latin{et~al.}(1985)Tomalia, Baker, Dewald, Hall, Kallos,
  Martin, Roeck, Ryder, and Smith]{tomalia_new_1985}
Tomalia,~D.~A.; Baker,~H.; Dewald,~J.; Hall,~M.; Kallos,~G.; Martin,~S.;
  Roeck,~J.; Ryder,~J.; Smith,~P. A {New} {Class} of {Polymers}:
  {Starburst}-{Dendritic} {Macromolecules}. \emph{Polym. J} \textbf{1985},
  \emph{17}, 117--132\relax
\mciteBstWouldAddEndPuncttrue
\mciteSetBstMidEndSepPunct{\mcitedefaultmidpunct}
{\mcitedefaultendpunct}{\mcitedefaultseppunct}\relax
\EndOfBibitem
\bibitem[Seifert(1997)]{seifert_configurations_1997}
Seifert,~U. Configurations of fluid membranes and vesicles. \emph{Adv. Phys.}
  \textbf{1997}, \emph{46}, 13--137\relax
\mciteBstWouldAddEndPuncttrue
\mciteSetBstMidEndSepPunct{\mcitedefaultmidpunct}
{\mcitedefaultendpunct}{\mcitedefaultseppunct}\relax
\EndOfBibitem
\bibitem[Banani \latin{et~al.}(2017)Banani, Lee, Hyman, and
  Rosen]{banani_biomolecular_2017}
Banani,~S.~F.; Lee,~H.~O.; Hyman,~A.~A.; Rosen,~M.~K. Biomolecular condensates:
  organizers of cellular biochemistry. \emph{Nat. Rev. Mol. Cell Biol.}
  \textbf{2017}, \emph{18}, 285--298\relax
\mciteBstWouldAddEndPuncttrue
\mciteSetBstMidEndSepPunct{\mcitedefaultmidpunct}
{\mcitedefaultendpunct}{\mcitedefaultseppunct}\relax
\EndOfBibitem
\bibitem[Gnan \latin{et~al.}(2017)Gnan, Rovigatti, Bergman, and
  Zaccarelli]{gnan_silico_2017}
Gnan,~N.; Rovigatti,~L.; Bergman,~M.; Zaccarelli,~E. In {Silico} {Synthesis} of
  {Microgel} {Particles}. \emph{Macromolecules} \textbf{2017}, \emph{50},
  8777--8786\relax
\mciteBstWouldAddEndPuncttrue
\mciteSetBstMidEndSepPunct{\mcitedefaultmidpunct}
{\mcitedefaultendpunct}{\mcitedefaultseppunct}\relax
\EndOfBibitem
\bibitem[Hofzumahaus \latin{et~al.}(2021)Hofzumahaus, Strauch, and
  Schneider]{hofzumahaus_monte_2021}
Hofzumahaus,~C.; Strauch,~C.; Schneider,~S. Monte {Carlo} simulations of weak
  polyampholyte microgels: {pH}-dependence of conformation and ionization.
  \emph{Soft Matter} \textbf{2021}, \emph{17}, 6029--6043\relax
\mciteBstWouldAddEndPuncttrue
\mciteSetBstMidEndSepPunct{\mcitedefaultmidpunct}
{\mcitedefaultendpunct}{\mcitedefaultseppunct}\relax
\EndOfBibitem
\bibitem[Daoud and Cotton(1982)Daoud, and Cotton]{daoud_star_1982}
Daoud,~M.; Cotton,~J. Star shaped polymers : a model for the conformation and
  its concentration dependence. \emph{J. Phys. (France)} \textbf{1982},
  \emph{43}, 531--538\relax
\mciteBstWouldAddEndPuncttrue
\mciteSetBstMidEndSepPunct{\mcitedefaultmidpunct}
{\mcitedefaultendpunct}{\mcitedefaultseppunct}\relax
\EndOfBibitem
\bibitem[Yu \latin{et~al.}(2020)Yu, Dasgupta, Auth, and
  Gompper]{yu_osmotic_2020}
Yu,~Q.; Dasgupta,~S.; Auth,~T.; Gompper,~G. Osmotic
  {Concentration}-{Controlled} {Particle} {Uptake} and {Wrapping}-{Induced}
  {Lysis} of {Cells} and {Vesicles}. \emph{Nano Lett.} \textbf{2020},
  \emph{20}, 1662--1668\relax
\mciteBstWouldAddEndPuncttrue
\mciteSetBstMidEndSepPunct{\mcitedefaultmidpunct}
{\mcitedefaultendpunct}{\mcitedefaultseppunct}\relax
\EndOfBibitem
\bibitem[Dimova and Marques(2019)Dimova, and Marques]{dimova_giant_2019}
Dimova,~R., Marques,~C.~M., Eds. \emph{The {Giant} {Vesicle} {Book}}; CRC
  Press: Boca Raton, 2019\relax
\mciteBstWouldAddEndPuncttrue
\mciteSetBstMidEndSepPunct{\mcitedefaultmidpunct}
{\mcitedefaultendpunct}{\mcitedefaultseppunct}\relax
\EndOfBibitem
\bibitem[Kukura \latin{et~al.}(2009)Kukura, Ewers, Müller, Renn, Helenius, and
  Sandoghdar]{kukura_high-speed_2009}
Kukura,~P.; Ewers,~H.; Müller,~C.; Renn,~A.; Helenius,~A.; Sandoghdar,~V.
  High-speed nanoscopic tracking of the position and orientation of a single
  virus. \emph{Nat. Methods} \textbf{2009}, \emph{6}, 923--927\relax
\mciteBstWouldAddEndPuncttrue
\mciteSetBstMidEndSepPunct{\mcitedefaultmidpunct}
{\mcitedefaultendpunct}{\mcitedefaultseppunct}\relax
\EndOfBibitem
\bibitem[Sarfati and Dufresne(2016)Sarfati, and
  Dufresne]{sarfati_long-range_2016}
Sarfati,~R.; Dufresne,~E.~R. Long-range attraction of particles adhered to
  lipid vesicles. \emph{Phys. Rev. E} \textbf{2016}, \emph{94}, 012604\relax
\mciteBstWouldAddEndPuncttrue
\mciteSetBstMidEndSepPunct{\mcitedefaultmidpunct}
{\mcitedefaultendpunct}{\mcitedefaultseppunct}\relax
\EndOfBibitem
\bibitem[E. Debets \latin{et~al.}(2020)E. Debets, C. Janssen, and
  Šarić]{edebets_characterising_2020}
E. Debets,~V.; C. Janssen,~L.~M.; Šarić,~A. Characterising the diffusion of
  biological nanoparticles on fluid and cross-linked membranes. \emph{Soft
  Matter} \textbf{2020}, \emph{16}, 10628--10639\relax
\mciteBstWouldAddEndPuncttrue
\mciteSetBstMidEndSepPunct{\mcitedefaultmidpunct}
{\mcitedefaultendpunct}{\mcitedefaultseppunct}\relax
\EndOfBibitem
\bibitem[Auth and Gompper(2009)Auth, and Gompper]{auth_budding_2009}
Auth,~T.; Gompper,~G. Budding and vesiculation induced by conical membrane
  inclusions. \emph{Phys. Rev. E} \textbf{2009}, \emph{80}, 031901\relax
\mciteBstWouldAddEndPuncttrue
\mciteSetBstMidEndSepPunct{\mcitedefaultmidpunct}
{\mcitedefaultendpunct}{\mcitedefaultseppunct}\relax
\EndOfBibitem
\bibitem[Reynwar \latin{et~al.}(2007)Reynwar, Illya, Harmandaris, Müller,
  Kremer, and Deserno]{reynwar_aggregation_2007}
Reynwar,~B.~J.; Illya,~G.; Harmandaris,~V.~A.; Müller,~M.~M.; Kremer,~K.;
  Deserno,~M. Aggregation and vesiculation of membrane proteins by
  curvature-mediated interactions. \emph{Nature} \textbf{2007}, \emph{447},
  461--464\relax
\mciteBstWouldAddEndPuncttrue
\mciteSetBstMidEndSepPunct{\mcitedefaultmidpunct}
{\mcitedefaultendpunct}{\mcitedefaultseppunct}\relax
\EndOfBibitem
\bibitem[Šarić and Cacciuto(2012)Šarić, and Cacciuto]{saric_fluid_2012}
Šarić,~A.; Cacciuto,~A. Fluid {Membranes} {Can} {Drive} {Linear}
  {Aggregation} of {Adsorbed} {Spherical} {Nanoparticles}. \emph{Phys. Rev.
  Lett.} \textbf{2012}, \emph{108}, 118101\relax
\mciteBstWouldAddEndPuncttrue
\mciteSetBstMidEndSepPunct{\mcitedefaultmidpunct}
{\mcitedefaultendpunct}{\mcitedefaultseppunct}\relax
\EndOfBibitem
\bibitem[Idema and Kraft(2019)Idema, and Kraft]{idema2019interactions}
Idema,~T.; Kraft,~D.~J. Interactions between model inclusions on closed lipid
  bilayer membranes. \emph{Curr. Opin. Colloid Interface Sci.} \textbf{2019},
  \emph{40}, 58--69\relax
\mciteBstWouldAddEndPuncttrue
\mciteSetBstMidEndSepPunct{\mcitedefaultmidpunct}
{\mcitedefaultendpunct}{\mcitedefaultseppunct}\relax
\EndOfBibitem
\bibitem[van~der Wel \latin{et~al.}(2016)van~der Wel, Vahid, Šarić, Idema,
  Heinrich, and Kraft]{van_der_wel_lipid_2016}
van~der Wel,~C.; Vahid,~A.; Šarić,~A.; Idema,~T.; Heinrich,~D.; Kraft,~D.~J.
  Lipid membrane-mediated attraction between curvature inducing objects.
  \emph{Sci. Rep.} \textbf{2016}, \emph{6}\relax
\mciteBstWouldAddEndPuncttrue
\mciteSetBstMidEndSepPunct{\mcitedefaultmidpunct}
{\mcitedefaultendpunct}{\mcitedefaultseppunct}\relax
\EndOfBibitem
\bibitem[Šarić and Cacciuto(2012)Šarić, and Cacciuto]{saric_mechanism_2012}
Šarić,~A.; Cacciuto,~A. Mechanism of {Membrane} {Tube} {Formation} {Induced}
  by {Adhesive} {Nanocomponents}. \emph{Phys. Rev. Lett.} \textbf{2012},
  \emph{109}, 188101\relax
\mciteBstWouldAddEndPuncttrue
\mciteSetBstMidEndSepPunct{\mcitedefaultmidpunct}
{\mcitedefaultendpunct}{\mcitedefaultseppunct}\relax
\EndOfBibitem
\bibitem[Acosta-Gutiérrez \latin{et~al.}(2021)Acosta-Gutiérrez, Buckley, and
  Battaglia]{acosta-gutierrez_role_2021}
Acosta-Gutiérrez,~S.; Buckley,~J.; Battaglia,~G. The role of host cell glycans
  on virus infectivity: {The} {SARS}-{CoV}-2 case. \emph{bioRxiv}
  \textbf{2021}, 2021.05.08.443212\relax
\mciteBstWouldAddEndPuncttrue
\mciteSetBstMidEndSepPunct{\mcitedefaultmidpunct}
{\mcitedefaultendpunct}{\mcitedefaultseppunct}\relax
\EndOfBibitem
\bibitem[Bahrami \latin{et~al.}(2012)Bahrami, Lipowsky, and
  Weikl]{bahrami_tubulation_2012}
Bahrami,~A.~H.; Lipowsky,~R.; Weikl,~T.~R. Tubulation and {Aggregation} of
  {Spherical} {Nanoparticles} {Adsorbed} on {Vesicles}. \emph{Phys. Rev. Lett.}
  \textbf{2012}, \emph{109}, 188102\relax
\mciteBstWouldAddEndPuncttrue
\mciteSetBstMidEndSepPunct{\mcitedefaultmidpunct}
{\mcitedefaultendpunct}{\mcitedefaultseppunct}\relax
\EndOfBibitem
\bibitem[Santiana \latin{et~al.}(2018)Santiana, Ghosh, Ho, Rajasekaran, Du,
  Mutsafi, De~Jésus-Diaz, Sosnovtsev, Levenson, Parra, Takvorian, Cali, Bleck,
  Vlasova, Saif, Patton, Lopalco, Corcelli, Green, and
  Altan-Bonnet]{santiana_vesicle-cloaked_2018}
Santiana,~M. \latin{et~al.}  Vesicle-{Cloaked} {Virus} {Clusters} {Are}
  {Optimal} {Units} for {Inter}-organismal {Viral} {Transmission}. \emph{Cell
  Host \& Microbe} \textbf{2018}, \emph{24}, 208--220.e8\relax
\mciteBstWouldAddEndPuncttrue
\mciteSetBstMidEndSepPunct{\mcitedefaultmidpunct}
{\mcitedefaultendpunct}{\mcitedefaultseppunct}\relax
\EndOfBibitem
\bibitem[Sanjuán and Thoulouze(2019)Sanjuán, and Thoulouze]{sanjuan_why_2019}
Sanjuán,~R.; Thoulouze,~M.-I. Why viruses sometimes disperse in groups†.
  \emph{Virus Evol.} \textbf{2019}, \emph{5}, vez014\relax
\mciteBstWouldAddEndPuncttrue
\mciteSetBstMidEndSepPunct{\mcitedefaultmidpunct}
{\mcitedefaultendpunct}{\mcitedefaultseppunct}\relax
\EndOfBibitem
\bibitem[Mkrtchyan \latin{et~al.}(2010)Mkrtchyan, Ing, and
  Chen]{mkrtchyan_adhesion_2010}
Mkrtchyan,~S.; Ing,~C.; Chen,~J. Z.~Y. Adhesion of cylindrical colloids to the
  surface of a membrane. \emph{Phys. Rev. E} \textbf{2010}, \emph{81},
  011904\relax
\mciteBstWouldAddEndPuncttrue
\mciteSetBstMidEndSepPunct{\mcitedefaultmidpunct}
{\mcitedefaultendpunct}{\mcitedefaultseppunct}\relax
\EndOfBibitem
\bibitem[Weikl(2003)]{weikl_indirect_2003}
Weikl,~T.~R. Indirect interactions of membrane-adsorbed cylinders. \emph{Eur.
  Phys. J. E} \textbf{2003}, \emph{12}, 265--273\relax
\mciteBstWouldAddEndPuncttrue
\mciteSetBstMidEndSepPunct{\mcitedefaultmidpunct}
{\mcitedefaultendpunct}{\mcitedefaultseppunct}\relax
\EndOfBibitem
\bibitem[Deserno(2004)]{deserno_elastic_2004}
Deserno,~M. Elastic deformation of a fluid membrane upon colloid binding.
  \emph{Phys. Rev. E} \textbf{2004}, \emph{69}, 031903\relax
\mciteBstWouldAddEndPuncttrue
\mciteSetBstMidEndSepPunct{\mcitedefaultmidpunct}
{\mcitedefaultendpunct}{\mcitedefaultseppunct}\relax
\EndOfBibitem
\bibitem[Reynwar and Deserno(2011)Reynwar, and
  Deserno]{reynwar_membrane-mediated_2011}
Reynwar,~B.~J.; Deserno,~M. Membrane-mediated interactions between circular
  particles in the strongly curved regime. \emph{Soft Matter} \textbf{2011},
  \emph{7}, 8567\relax
\mciteBstWouldAddEndPuncttrue
\mciteSetBstMidEndSepPunct{\mcitedefaultmidpunct}
{\mcitedefaultendpunct}{\mcitedefaultseppunct}\relax
\EndOfBibitem
\bibitem[Kim \latin{et~al.}(1998)Kim, Neu, and
  Oster]{kim_curvature-mediated_1998}
Kim,~K.; Neu,~J.; Oster,~G. Curvature-{Mediated} {Interactions} {Between}
  {Membrane} {Proteins}. \emph{Biophys. J.} \textbf{1998}, \emph{75},
  2274--2291\relax
\mciteBstWouldAddEndPuncttrue
\mciteSetBstMidEndSepPunct{\mcitedefaultmidpunct}
{\mcitedefaultendpunct}{\mcitedefaultseppunct}\relax
\EndOfBibitem
\bibitem[Weikl \latin{et~al.}(1998)Weikl, Kozlov, and
  Helfrich]{weikl_interaction_1998}
Weikl,~T.~R.; Kozlov,~M.~M.; Helfrich,~W. Interaction of conical membrane
  inclusions: {Effect} of lateral tension. \emph{Phys. Rev. E} \textbf{1998},
  \emph{57}, 6988--6995\relax
\mciteBstWouldAddEndPuncttrue
\mciteSetBstMidEndSepPunct{\mcitedefaultmidpunct}
{\mcitedefaultendpunct}{\mcitedefaultseppunct}\relax
\EndOfBibitem
\bibitem[Goulian \latin{et~al.}(1993)Goulian, Bruinsma, and
  Pincus]{goulian_long-range_1993}
Goulian,~M.; Bruinsma,~R.; Pincus,~P. Long-{Range} {Forces} in {Heterogeneous}
  {Fluid} {Membranes}. \emph{Europhys. Lett.} \textbf{1993}, \emph{22},
  145--150\relax
\mciteBstWouldAddEndPuncttrue
\mciteSetBstMidEndSepPunct{\mcitedefaultmidpunct}
{\mcitedefaultendpunct}{\mcitedefaultseppunct}\relax
\EndOfBibitem
\bibitem[Baumgart \latin{et~al.}(2003)Baumgart, Hess, and
  Webb]{baumgart2003imaging}
Baumgart,~T.; Hess,~S.~T.; Webb,~W.~W. Imaging coexisting fluid domains in
  biomembrane models coupling curvature and line tension. \emph{Nature}
  \textbf{2003}, \emph{425}, 821--824\relax
\mciteBstWouldAddEndPuncttrue
\mciteSetBstMidEndSepPunct{\mcitedefaultmidpunct}
{\mcitedefaultendpunct}{\mcitedefaultseppunct}\relax
\EndOfBibitem
\bibitem[Noguchi and Fournier(2017)Noguchi, and
  Fournier]{noguchi_membrane_2017}
Noguchi,~H.; Fournier,~J.-B. Membrane structure formation induced by two types
  of banana-shaped proteins. \emph{Soft Matter} \textbf{2017}, \emph{13},
  4099--4111\relax
\mciteBstWouldAddEndPuncttrue
\mciteSetBstMidEndSepPunct{\mcitedefaultmidpunct}
{\mcitedefaultendpunct}{\mcitedefaultseppunct}\relax
\EndOfBibitem
\bibitem[Scotti \latin{et~al.}(2022)Scotti, Schulte, Lopez, Crassous, Bochenek,
  and Richtering]{scotti2206}
Scotti,~A.; Schulte,~F.; Lopez,~C.~G.; Crassous,~J.~J.; Bochenek,~S.;
  Richtering,~W. How Softness Matters in Soft Nanogels and Nanogel Assemblies.
  \emph{Chem. Rev.} \textbf{2022}, \relax
\mciteBstWouldAddEndPunctfalse
\mciteSetBstMidEndSepPunct{\mcitedefaultmidpunct}
{}{\mcitedefaultseppunct}\relax
\EndOfBibitem
\bibitem[Gompper and Kroll(2004)Gompper, and
  Kroll]{gompper_triangulated-surface_2004}
Gompper,~G.; Kroll,~D.~M. \emph{Statistical {Mechanics} of {Membranes} and
  {Surfaces}}; WORLD SCIENTIFIC, 2004; pp 359--426\relax
\mciteBstWouldAddEndPuncttrue
\mciteSetBstMidEndSepPunct{\mcitedefaultmidpunct}
{\mcitedefaultendpunct}{\mcitedefaultseppunct}\relax
\EndOfBibitem
\bibitem[Sunil~Kumar \latin{et~al.}(2001)Sunil~Kumar, Gompper, and
  Lipowsky]{sunil_kumar_budding_2001}
Sunil~Kumar,~P.~B.; Gompper,~G.; Lipowsky,~R. Budding {Dynamics} of
  {Multicomponent} {Membranes}. \emph{Phys. Rev. Lett.} \textbf{2001},
  \emph{86}, 3911--3914\relax
\mciteBstWouldAddEndPuncttrue
\mciteSetBstMidEndSepPunct{\mcitedefaultmidpunct}
{\mcitedefaultendpunct}{\mcitedefaultseppunct}\relax
\EndOfBibitem
\bibitem[Kroll and Gompper(1992)Kroll, and Gompper]{kroll_conformation_1992}
Kroll,~D.~M.; Gompper,~G. The {Conformation} of {Fluid} {Membranes}: {Monte}
  {Carlo} {Simulations}. \emph{Science} \textbf{1992}, \emph{255},
  968--971\relax
\mciteBstWouldAddEndPuncttrue
\mciteSetBstMidEndSepPunct{\mcitedefaultmidpunct}
{\mcitedefaultendpunct}{\mcitedefaultseppunct}\relax
\EndOfBibitem
\bibitem[Brakke(1992)]{brakke_surface_1992}
Brakke,~K.~A. The {Surface} {Evolver}. \emph{Exp. Math.} \textbf{1992},
  \emph{1}, 141--165\relax
\mciteBstWouldAddEndPuncttrue
\mciteSetBstMidEndSepPunct{\mcitedefaultmidpunct}
{\mcitedefaultendpunct}{\mcitedefaultseppunct}\relax
\EndOfBibitem
\end{mcitethebibliography}

\end{document}